\def\HI{H{\,\small I}}
\documentclass{aa}  
\usepackage{graphicx}
\usepackage[varg]{txfonts}
\usepackage[pdftex, colorlinks=true,linkcolor=blue,citecolor=blue]{hyperref}
\usepackage{natbib}

\usepackage{xspace}
\usepackage{multirow}

\usepackage[mathlines,switch]{lineno}

\begin{document} 

   \title{The parsec-scale HI outflows in powerful radio galaxies}

   \author{R. Schulz\inst{1}
          \and
          R. Morganti\inst{1,2}
          \and
          K. Nyland\inst{3}
          \and
			Z. Paragi\inst{4}
			\and
			E. K. Mahony\inst{5}
          \and
          T. Oosterloo\inst{1,2}
          }

   \institute{ASTRON, Netherlands Institute for Radio Astronomy, Oude Hoogeveensedijk 4, 7991 PD, Dwingeloo, Netherlands\\
			 \email{r.f.schulz@issc.leidenuniv.nl,morganti@astron.nl}
			 \and
			 Kapteyn Astronomical Institute, University of Groningen, PO Box 800, 9700 AV Groningen, The Netherlands 
			\and 
			National Research Council, resident at the U.S. Naval Research Laboratory, 4555 Overlook Ave. SW, Washington, DC 20375, USA
			\and
			Joint Institute for VLBI ERIC, Oude Hoogeveensedijk 4, 7991 PD Dwingeloo, The Netherlands
			\and
			Australia Telescope National Facility, CSIRO Astronomy and Space Science, PO Box 76, Epping, NW 1710, Australia
             }

   \date{--; --}

  \abstract
  {
		Massive outflows of neutral atomic hydrogen (\ion{H}{I}) have been observed in absorption in a number of radio galaxies and are considered a signature of AGN feedback. These outflows on kpc-scales have not been investigated in great detail as they require high-angular resolution observations to be spatially resolved. In some radio AGN, they are likely the result of the radio jets interacting with the interstellar medium.	We have used the global very-long-baseline-interferometry (VLBI) array to map the \ion{H}{I} outflow in a small sample of young and restarted radio galaxies which we previously observed with the Very Large Array (VLA) and the Westerbork Synthesis Radio Telescope (WSRT) at lower resolution.  Here, we report on our findings for \object{4C\,52.37} and \object{3C\,293} and we discuss the sample including the previously published \object{4C\,12.50} and \object{3C\,236}. 
		For \object{4C\,52.37}, we present the first-ever \ion{H}{I} VLBI observations which recovered the majority of the outflowing \ion{H}{I} gas in form of clouds towards the central 100\,pc of the AGN. The clouds are blue-shifted by up to $\sim 600 ~\mathrm{km\,s^{-1}}$ with respect to the systemic velocity. \object{3C\,293} is largely resolved out in our VLBI observation, but we detect, towards the VLBI core, some outflowing \ion{H}{I} gas blueshifted with respect to the systemic velocity by up to $\sim 300~\mathrm{km\,s^{-1}}$. We also find indications of outflowing gas towards the other parts of the western lobe  suggesting that the \HI\ outflow is extended.
		
		Overall, we find that the fraction of \ion{H}{I} gas recovered by our VLBI observation varies significantly within our sample ranging from complete (\object{4C\,12.50}) to marginal (\object{3C\,293}). However, in all cases we find evidence for a clumpy structure of both the outflowing and the quiescent gas, consistent with predictions from numerical simulations. All the outflows include at least a component of relatively compact clouds with masses in the range of $10^4-10^5M_\sun$. The outflowing clouds are often observed already at a few tens of pc (in projection) from the core. We find indications that the \ion{H}{I} outflow might have a diffuse component, especially in larger sources. Our results support the interpretation that we observe these AGNs at different stages in the evolution of the jet-ISM interaction and this is reflected in the properties of the outflowing gas as predicted by numerical simulations.
  }

   \keywords{Galaxies: active - Galaxies: jets - Galaxies: individual: \object{4C\,12.50}, \object{4C\,52.37}, \object{3C\,236}, \object{3C\,295} - Galaxies: nuclei - Techniques: high angular resolution - ISM: jets and outflows}
   
   \maketitle
%
\section{Introduction}
\label{sec:Intro}

	Active galactic nuclei (AGNs) produce tremendous amounts of energy through the accretion of matter onto the super-massive black hole (SMBH). The release of this energy can heat and expel gas that would otherwise be available for star formation and for the growth of the SMBH. Thus, the energy release can directly impact the interstellar medium (ISM) of the host galaxy and this interplay, or feedback, links the evolution of the galaxy to the activity of the SMBH  (e.g., \citealt{Silk1998,DiMatteo2005,Croton2006,McNamara2007,Heckman2014,Harrison2017}). 

	Among the clear observational evidence of this interaction are outflows of gaseous matter observed in all phases of the ISM. Outflows of hot and warm ionized gas have been found in many AGNs over a range of redshifts suggesting that they could be a general feature (e.g., \citealt{Veilleux2005,Tadhunter2008,King2015,Morganti2017b}).

	In addition, several studies have shown outflows of matter from the cold ISM, i.e., molecular and atomic gas. Interestingly, the outflows of cold gas appear to carry higher mass outflow rates than their counterparts of ionized gas (e.g., \citealt{Morganti2005,Cicone2014,Morganti2017b,Fiore2017,Veilleux2020}).
	There are different possible driving mechanisms capable of producing these outflows, i.e. radiative winds from the accretion disk, radiation pressure on dust, and radio jets. Observational evidence has been found for all of them with one being more likely than other depending on the type of host galaxy and nuclear activity mode (e.g., \citealt{McNamara2012,King2015,Morganti2017b}). 

	In a number of cases, radio jets appear to be the most likely mechanism capable to accelerate the gas and produce outflows (e.g., \citealt{Morganti1998,Oosterloo2000,Morganti2005,Morganti2005b,Morganti2013,Morganti2016,Mahony2013,Gereb2015,Aditya2018,Aditya2018c}).
	In particular, one of the findings supporting this is the higher fraction of outflows observed in young and restarted AGNs and, in some cases, the close morphological association of the region of the outflow and the radio jet (e.g., \citealt{Holt2008}).  

	Radio galaxies in their young phase are often characterised as Compact Steep Spectrum (CSS) or GHz-peaked spectrum (GPS) sources based on the shape of the radio spectrum (e.g., \citealt{ODea1998,Orienti2016}).
	The radio continuum emission of these sources extends to a few kpc or less and it is still contained within the host galaxy. They are often hosted in galaxies with a richer ISM \citep{Callingham2017,Holt2009} which has also been confirmed by the study of \ion{H}{I} gas \citep{Gereb2015,Gupta2006,Chandola2011}.

	In restarted AGNs, the CSS/GPS source stems from the most recent or current cycle of AGN activity and is embedded in larger radio galaxies where the large-scale radio emission reflects previous cycles of AGN activity.
	The expected coupling of the jet with the ISM depends on the properties of the gas. In particular, in recent numerical simulations (\mbox{\citealt{Wagner2011}}; \mbox{\citealt{Wagner2012}}; \mbox{\citealt{Mukherjee2016,Mukherjee2018}}; \mbox{\citealt{Cielo2017b}}) 
	the ISM is now modeled much more realistically (e.g. clumpy instead of smooth). An important result is that a clumpy ISM has a much stronger impact on the propagation of newly (re-)born jets: the jet is interacting with dense clumps of the ISM, temporarily blocking its progress, which results in an over-pressured cocoon in the ISM which affects a much larger region of the host galaxy than the jet itself. 
	Investigating whether this scenario is confirmed by the observations requires to reach high spatial resolution to resolve the distribution and kinematics of the outflowing gas and compare it with the predictions of the simulations. Such a comparison has been already successful in some cases, e.g.  \cite{Mukherjee2018,Zovaro2019}, but the number is still limited. 

	In this work, we focus on deriving the properties and structure of outflows of neutral atomic hydrogen (\ion{H}{I})
	gas observed in absorption in young or restarted radio galaxies. In order to resolve the outflows on sub-arcsecond angular resolution we carried out Very Long Baseline Interferometry (VLBI) observations. This paper is a continuation of the work presented in \cite{Morganti2013} on \object{4C\,12.50} and in \cite{Schulz2018} on \object{3C\,236}, both of which are restarted radio sources. Here, we present results from the remaining two sources of our initial sample \object{4C\,52.37} and \object{3C\,293} and compare the properties of the \ion{H}{I} outflows in all four sources in particular with respect to the evolution of the radio AGNs.

	The paper is structured as follows: we first describe the sample and in particular the properties of \object{4C\,52.37} and \object{3C\,293}. This is followed by a description of the VLBI observation and data reduction in Sect. \ref{sec:Obs}. This is followed by the presentation of our results on the line and continuum in Sect. \ref{sec:Results}. We discuss in Sect. \ref{sec:Discussion} the results of the two new objects and we also discuss some general results obtained by combining the findings for the full sample. We conclude this paper with a summary in Sect. \ref{sec:Summary} and provide additional plots in Appendix \ref{sec:appendix}. Throughout this paper, we use a $\Lambda$CDM cosmology of $H_0=70\mathrm{\,km\,s^{-1}\,Mpc^{-1}}$, $\Omega_m = 0.3$ and $\Omega_\Lambda=0.7$. 

\section{The sample}
\label{sec:sample}

	The four sources \object{4C\,12.50}, \object{3C\,236}, \object{4C\,52.37}, and \object{3C\,293} have been selected because they exhibit fast \ion{H}{I}
	outflows, they are powerful radio sources and they are likely to be young or restarted AGNs, but they are potentially at different stages in their AGN evolution. Based on their optical properties, \object{4C\,52.37}, \object{3C\,236}, and \object{3C\,293} have been classified as low-excitation radio galaxies \citep{Buttiglione2010,Best2012,deGasperin2011} which makes the radio jet the most likely driving force behind the outflowing gas. Even though \object{4C\,12.50} is classified as a high-excitation radio galaxy \citep{Holt2011}, \cite{Morganti2013} concluded that the \ion{H}{I} outflow is jet-driven based on the location of the \ion{H}{I} outflow. The VLBI observations revealed all of the \ion{H}{I} outflow concentrated in a slightly extended cloud towards the hot spot of the southern radio jet. In \object{3C\,236}, part of the \ion{H}{I} outflow was detected in the form of unresolved clouds towards the center of the AGN and as an extended cloud located towards the hot spot of the south-east radio lobe, but the VLBI data did not recover all of the outflow and suggested that part of the outflow might be in the form of a diffuse component. 	

	\object{4C\,52.37} is a radio galaxy located at a redshift of 0.106 based on SDSS \citep{Abolfathi2017}. The SDSS redshift corresponds to a systemic optical velocity of $31688\mathrm{\,km\,s^{-1}}$ and a linear scale of $1\mathrm{\,mas} \approx 1.9\mathrm{\,pc}$. The optical spectrum is characterised by strong broad permitted and forbidden emission lines. \object{4C\,52.37} is included in CORALZ sample of compact and young radio sources \citep{Snellen2004}. The sub-arcsecond radio morphology has been classified as a compact symmetric object by previous VLBI observations (\citealt{deVries2009}). So far, it is not clear whether there have been previous cycles of radio activity. The main \ion{H}{I} absorption feature was first reported by \cite{Chandola2011} using the Giant Metrewave Radio Telescope (GMRT). The \ion{H}{I} outflow was discovered by observations with the Westerbork Synthesis Radio Telescope (WSRT, \citealt{Gereb2015,Maccagni2017}) and covers more than $600\mathrm{\,km\,s^{-1}}$. 
	For \object{4C\,52.37}, this study represents the first detailed investigation of the \ion{H}{I} absorption with VLBI.

	The radio galaxy \object{3C\,293} is located at a redshift of 0.045 \citep{Abolfathi2017}. At this redshift an angular scale of 1 mas corresponds to 0.96 pc. Optical observations show a complex system of dust lanes and a companion galaxy \citep{Heckman1986,Martel1999,Evans1999}. The host galaxy of \object{3C\,293} has undergone at least one merger event, but it is unclear whether the companion galaxy is connected to this \citep{Evans1999,Capetti2002,Floyd2006}. The star formation and radio activity is considered to be linked to the merger event (\citealt{Tadhunter2005,Labiano2014}). The large-scale radio emission extends over 2\arcmin{} with a bright central region comprising of two strong features extending about 2\arcsec{} in size and oriented in an east-west direction (e.g., \citealt{Haschick1985,Beswick2002}). Strong \ion{H}{I} absorption has been detected from both features of the central region \citep{Baan1981,Haschick1985,Beswick2002} and a shallow, broad ($\sim1200\mathrm{\,km\,s^{-1}}$) blue wing corresponding to an \ion{H}{I} outflow towards the western feature \citep{Morganti2003,Morganti2005,Mahony2013}. VLBI observations by \cite{Beswick2004} reveal a complex, extended radio morphology of the central region with \ion{H}{I} absorption corresponding to the dust lane detected over most of the radio emission. However, the available bandwidth of the observations by \cite{Beswick2004} was insufficient to detect the outflow. Cold molecular gas in the form of CO has been detected in absorption and emission \citep{Evans1999,Labiano2014} and associated with an asymmetric, warped disk rotating around the AGN. \cite{Labiano2014} did not detect a high-velocity molecular outflow associated with \object{3C\,293} based on their CO observation.  
	
\section{Observation \& Data reduction}
\label{sec:Obs}
	\begin{table*}
		\centering
		\caption[]{Properties of Observations}
		\label{tab:Data:Observation}
		\setlength{\tabcolsep}{4pt}
		\begin{tabular}{ccccccccccccc}
			\hline
			Source & $z$ & Array\tablefootmark{a} & Code\tablefootmark{b} & Date & $\nu_\mathrm{obs}$\tablefootmark{c} & $T_\mathrm{obs}$\tablefootmark{d} & Pol.\tablefootmark{e} & Correlator Pass & IFs & BW\tablefootmark{f} & $N_\mathrm{ch}$\tablefootmark{g} & $\Delta \nu$\tablefootmark{h}\\
			& & & & [GHz] & [min] & & & & [MHz] & & [kHz] \\
			\hline
			\hline
			\object{\mbox{3C\,293}}& 0.045 & EVN+VLBA+Ar & GN002A & 2015-03-02 & 1.36 & 500 & Dual & continuum & 4 & 16 & 32 & 500\\
			&	&   &	&	&	&	&	& spectral-line & 1 & 16 & 512 & 31.25\\
			\object{\mbox{4C\,52.37}}& 0.106 & EVN+VLBA+Ar & GN002C & 2015-10-15 & 1.28 & 500  & Dual & continuum & 4 & 16 & 32 & 500\\
			&	&  &	&	&	&	&	& spectral-line & 1 & 16 & 512 & 31.25\\
			\hline
		\end{tabular}
		\tablefoot{ 
			\tablefoottext{a}{Array used for observation. EVN: Effelsberg (Germany), phased-up Westerbork (5 stations, Netherlands), Jodrell-Bank (United Kingdom), Onsala (Sweden); VLBA (USA): Los Alamos (NM), Kitt Peak (AZ), St. Croix (VI), Mauna Kea (HI), Hancock (NH), Brewster (WA), Fort Davis (TX), North Liberty (IA), Pie Town (NM), Owens Valley (CA); Ar: Arecibo (Puerto Rico).}
			\tablefoottext{b}{Experiment Code}
			\tablefoottext{c}{Observing frequency}
			\tablefoottext{e}{Polarization: Dual refers to two polarization were used (LL and RR)}
			\tablefoottext{d}{Observing time. For the VLBI experiment, this represents the total on source time of the whole array.}
			\tablefoottext{f}{Bandwidth (of each IF)}
			\tablefoottext{g}{Number of channels in a single band or IF}
			\tablefoottext{h}{Channel width in frequency}
		}
	\end{table*}

	\begin{table*}[htpb]
		\centering
		\caption[]{Properties of Images}
		\label{tab:Data:Image}
		\begin{tabular}{cccccccc}
			\hline
			Source & Data & $\sigma_\mathrm{noise}$\tablefootmark{a} & Beam\tablefootmark{b} & $\Delta v$\tablefootmark{c} & $N_\mathrm{ch}$\tablefootmark{d} & $S_\mathrm{peak}$\tablefootmark{e} & $S_\mathrm{tot}$\tablefootmark{f}\\
			 & & [mJy\,beam$^{-1}$\,ch$^{-1}$] & [mas $\times$ mas] & [km\,s$^{-1}$] & & [Jy\,beam$^{-1}$] & [Jy] \\
			\hline
			\hline
			\object{\mbox{3C\,293}} & Continuum    & 0.19 & $25\times 25$ & & & 0.066 & 0.539 \\
			 & Cube 	     & 0.31 & $25\times 25$ & 20.7 (41.4) & 143 & &\\
			\object{\mbox{4C\,52.37}} & Continuum     & 0.19 & $6\times 6$ & & & 0.025 & 0.429 \\
			 & Continuum    & 0.15  & $20\times 20$ &    &   & 0.051 & 0.429\\
			 & Cube 	     & 0.58 & $20\times 20$ & 21.9 (43.8) & 143 & &\\
			 \hline
		\end{tabular}
		\tablefoot{ 
			\tablefoottext{a}{Noise level. For the cubes, this value represents the average value over all channels}
			\tablefoottext{b}{Synthesized beam as major axis and minor axis}
			\tablefoottext{c}{Channel width in velocity of the cubes. The values in brackets represent the effective resolution because of Hanning smoothing}
			\tablefoottext{d}{Number of channels of the cubes}
			\tablefoottext{f}{Peak brightness of the continuum image}
			\tablefoottext{e}{Total flux density of the continuum image}
		}
	\end{table*}

	\subsection{4C 52.37}
	\label{sec:Obs:4C52}

		\object{\mbox{4C\,52.37}} was observed with a global VLBI array comprising the European VLBI Network (EVN), the Very Long Baseline Array (VLBA) and the Arecibo radio telescope on 2015 Oct 15 (project code: GN002C). The data were correlated at JIVE producing two data sets: a 'continuum pass` with 4 IFs each with 32 channels and 16\,MHz bandwidth and a 'spectral-line pass` with 1 IF of 16\,MHz bandwidth and 512 channels (see Table \ref{tab:Data:Observation}). We used \object{J1545$+$5400} and \object{J1638+5720} as the phase reference source and fringe finder/bandpass calibrator, respectively.

		The data reduction process is similar to \cite{Schulz2018}. 
		We processed both the continuum and spectral-line pass in AIPS (version  31DEC16, \citealt{AIPS1999}), using the a-priori amplitude calibration and flag tables provided by the EVN calibration pipeline. We first started with the calibration of the continuum pass. We flagged the data if the station elevation was below 15\degr. Next, we corrected for the instrumental delay using the task FRING and \object{J1638+5720}. Due to the nature of the VLBI array, a single scan of \object{J1638+5720} from all stations was not available. Therefore, we used two scans for the manual phase calibration. In the next step, we corrected for the phase, delay and rate of both calibrators using a global fringe fit. The solutions for each calibrator were applied to themselves. In addition, the solution of the phase calibrator  \object{J1545$+$5400} was applied to our target source (\object{4C\,52.37}). Finally, we performed the bandpass calibration using \object{J1638+5720} and applied the solutions to all sources. Because the global fringe fit assumes a point source model, we imaged the calibrators and found them to be slightly extended. Therefore, we used the image of the calibrators as an input model for the global fringe fit and repeated the process up to the bandpass calibration.

		For the spectral-line pass, we used the amplitude calibration provided by the EVN calibration pipeline and the calibration of the continuum pass. We first applied the solutions from the manual phase calibration of the continuum pass to the spectral-line pass. Afterwards, we took the solutions of the calibration from the continuum pass and applied them to the calibrator and target source of the spectral-line pass. In the last step, we performed the bandpass calibration. Afterwards, we produced a continuum and a spectral-line visibility data file of \object{4C\,52.37}. For the continuum data, we averaged all channels together, while the spectra-line data contains the full cube.

		Flagging bad visibilities, imaging and self-calibration of the continuum data of \object{4C\,52.37} was performed with DIFMAP \citep{Shepherd1994,Difmap2011}. We used natural weighting and set CLEAN windows to allow for better convergence of the CLEAN algorithm. Once a sufficiently good model was achieved, we performed phase self-calibration and imaged the data again. This process was repeated with smaller time intervals for the phase self-calibration as long as the solutions stayed smooth. Then, a constant gain correction factor over the entire observing time was determined with amplitude self-calibration and applied to the data. After flagging bad solutions, we repeated the smaller cycle of imaging and phase-self-calibration and performed another amplitude self-calibration with a smaller time interval. We repeated this process with decreasing time intervals for amplitude self-calibration. The noise level of the final continuum image is above the theoretical thermal noise level which is expected in phase referencing VLBI. Also, radio frequency interference in this frequency band can contribute to an increase in noise level.
		
    	For further processing of the spectral-line visibility data of \object{4C\,52.37}, we first inspected the data in detail and flagged bad visibilities. In order to keep the visibilities as uniform as possible across channels, we generally flagged entire scans or channels of a specific telescope instead of single visibilities. Next, we loaded the continuum image into AIPS and used it to perform a single phase self-calibration of the spectral-line visibility data. To improve sensitivity and consistency with the continuum data, we time averaged data within 30s. In the next step, we subtracted the continuum in the visibility domain using a fit to the first and last 100 channels. 
    	Because we are primarily interested in the broad absorption and in order to increase the sensitivity, we averaged the spectra-line data also in frequency. This reduced the velocity resolution by a factor of 3 to $20.7\mathrm{\,km\,s^{-1}}$. The resulting data cube was manually imaged in AIPS using robust weighting 1, a visibility taper of 0.3 at $10\mathrm{M\lambda}$ and a restoring beam of 20\,mas. 
    	Like in \cite{Schulz2018} we found this combination of parameters to be sufficient. We cleaned only those channels of the cube that showed significant absorption in the area of the radio continuum. 
    	The imaged cube was Hanning-smoothed to further improve the sensitivity which reduces the effective velocity resolution by an additional factor of 2. The cube was de-redshifted in order to gain relative velocity information in the source rest frame.
    	The full list of image parameters is given in Table \ref{tab:Data:Observation}.

	\subsection{3C 293}
	\label{sec:Obs:3C293}

		\object{\mbox{3C\,293}} was observed with a global VLBI array on 2015 March 2 (project code: GN002A). \object{\mbox{J1350$+$3034}} and \object{\mbox{J1407$+$2827}} served as phase reference source and bandpass calibrator, respectively. Correlation and data processing was done following the procedure outlined for \object{\mbox{4C\,52.37}}.

		Unfortunately, an issue in the frequency setup of the observation resulted in a shift of the center frequency. As a result a significant part of the blue-shifted wing observed in previous observations was shifted out of the band. Therefore, only a linear fit using the outer 200 channels
		on the high-velocity side of the spectrum could be used
 		for the subtraction of the continuum in the visibility domain. We used a restoring beam of 25\,mas for the continuum image and spectra-line cube instead of the highest angular resolution due to the large extent of the radio emission.

	\subsection{Archival radio data}
	\label{sec:Obs:archival}

		We make use of archival Very Large Array (VLA) and WSRT data for all four sources in this paper. The VLA data of \object{3C\,236} and \object{3C\,293} were presented in \mbox{\cite{Schulz2018}} and \mbox{\cite{Mahony2013}}, while the WSRT data on \object{4C\,12.50} and \object{4C\,52.37} were taken from \mbox{\cite{Morganti2013}} and \mbox{\cite{Gereb2015}}. We also made use of VLBI data from \object{4C\,12.50} \mbox{\citep{Morganti2013}} and \object{3C\,236} \mbox{\citep{Schulz2018}}.
	
		In Appendix \ref{sec:appendix} we present the \ion{H}{I} profiles of these objects, which have been fitted with Gaussian distribution (using the Python package LMFIT \citealt{Newville2014}) to characterize the absorption spectrum.  The fit parameters are listed in Table \ref{tab:Data:fitparam} and plots of the fits are shown in Figs. \ref{fig:4C12_spec_fit} to \ref{fig:3C293_spec_fit}. These parameters are slightly different from \cite{Maccagni2017} because that study used the busy function \citep{Westmeier2014} to fit the spectrum.  In order to be consistent, we use Gaussian distribution for all our objects for an easy comparison. 
		
		\begin{figure}
			\centering
			\includegraphics[width=0.97\linewidth]{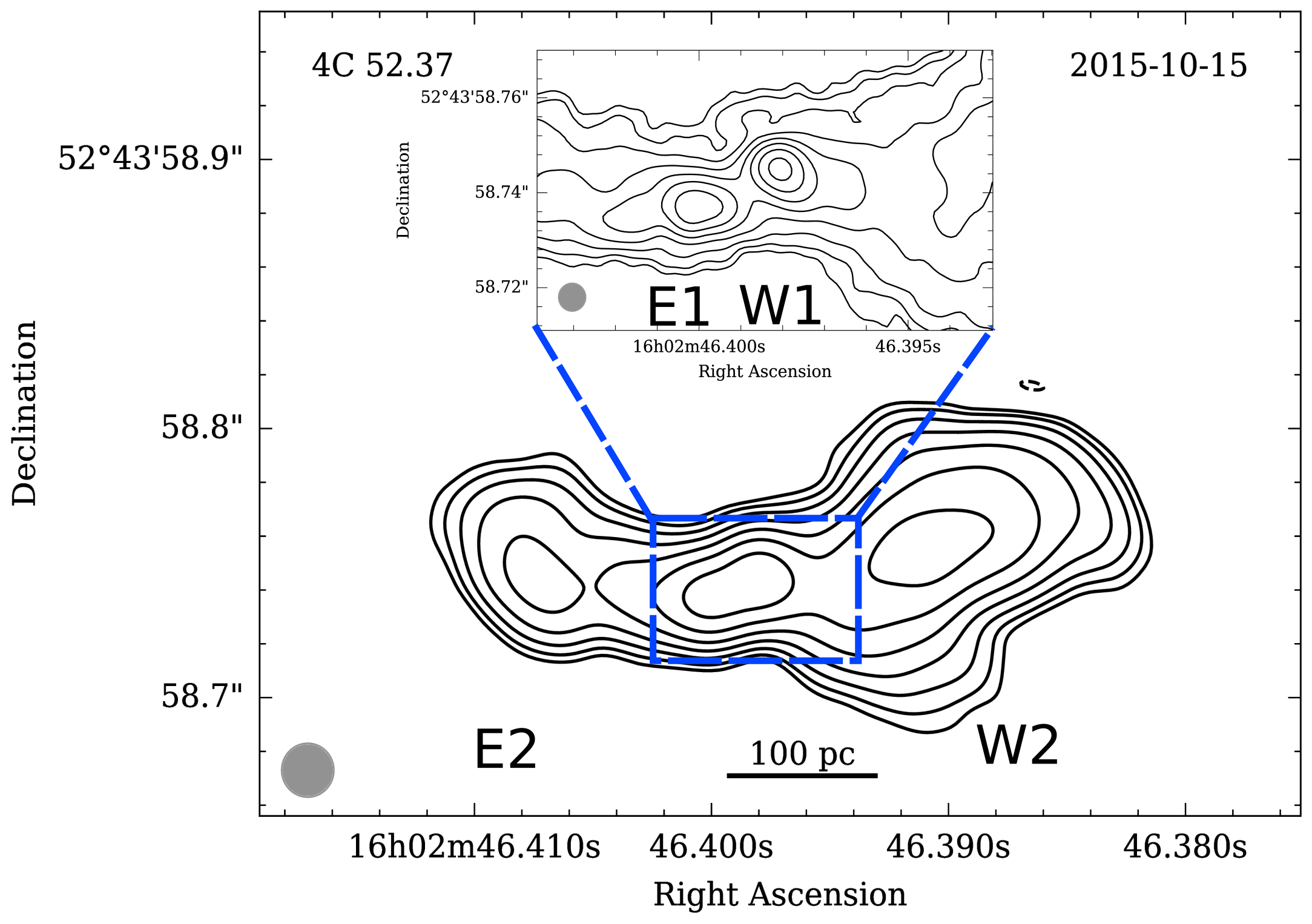}
			\includegraphics[width=0.97\linewidth]{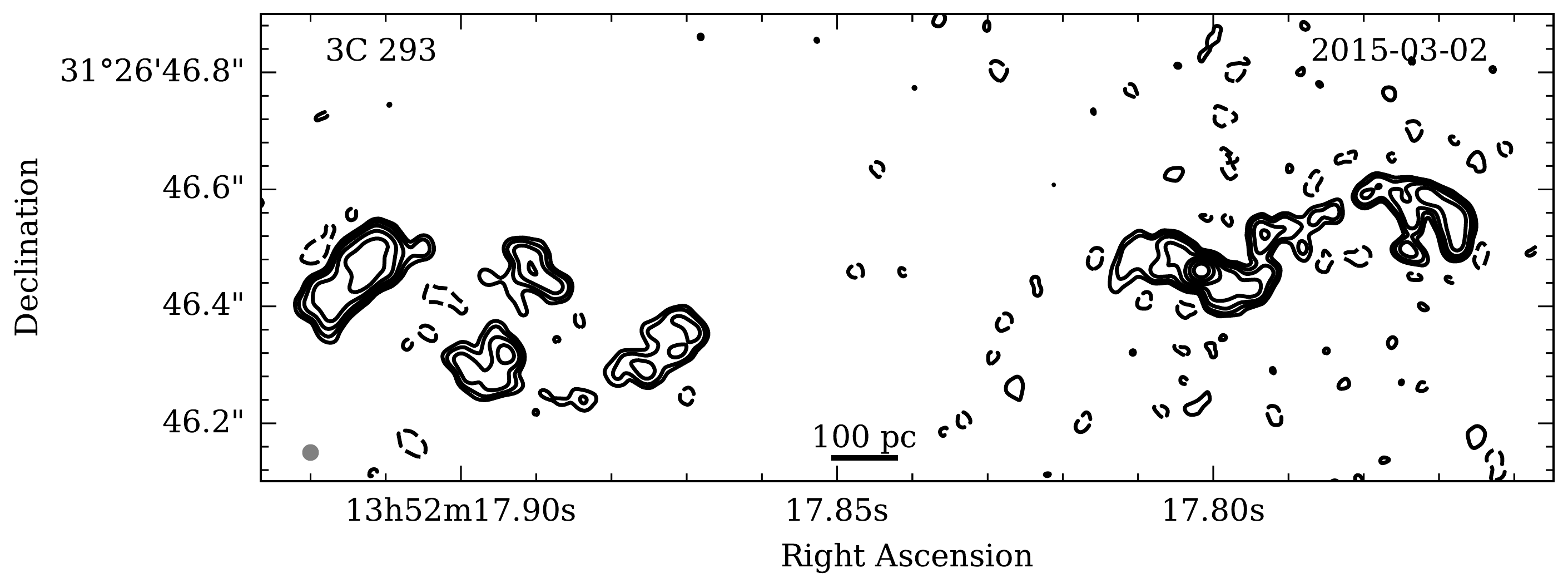}
			\caption{Continuum images of 4C52.37 (top) and \object{3C\,293} (bottom) convolved with a synthesized beam of 20\,mas and 25\,mas, respectively, which is marked by the gray circle. The top panel also shows a higher-resolution (6\,mas) zoom-in of the central region of the radio source. The contour lines begin at 3 times the noise level of the corresponding image and increase logarithmically by a factor of 2. The solid, horizontal scale bar represents the spatial scales.}
			\label{fig:4C52_3C293_cont}%
		\end{figure}

		\begin{figure}
			\includegraphics[width=0.97\linewidth]{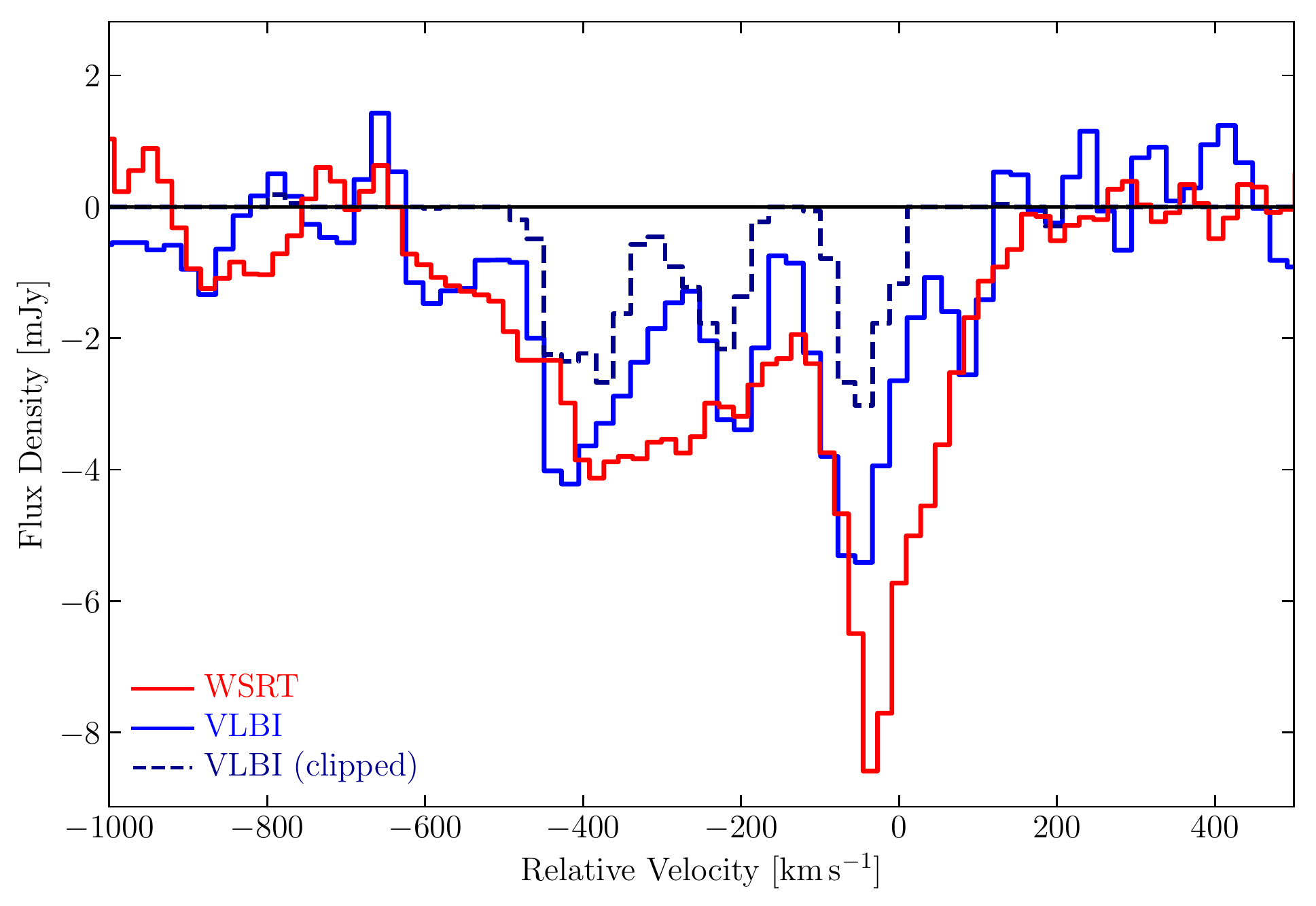}
			\centering
			\caption{\ion{H}{I} absorption spectra of 4C52.37. The unresolved WSRT spectrum is shown in red. The two spatially integrated VLBI profiles (in blue) were extracted from a region of $60\mathrm{\,mas}\times 40\mathrm{\,mas}$ centered on E1. The dashed blue line corresponds to the VLBI spectrum that was obtained by integrating only pixels with a value of $\leq-3\sigma_\mathrm{VLBI,cube}$, while for the solid blue line the integration was only limited to the region marked by $3\sigma_\mathrm{VLBI,cont}$ contour.}
			\label{fig:4C52_spec}%
		\end{figure}

		\begin{figure}
			\centering
			\includegraphics[width=0.9\linewidth]{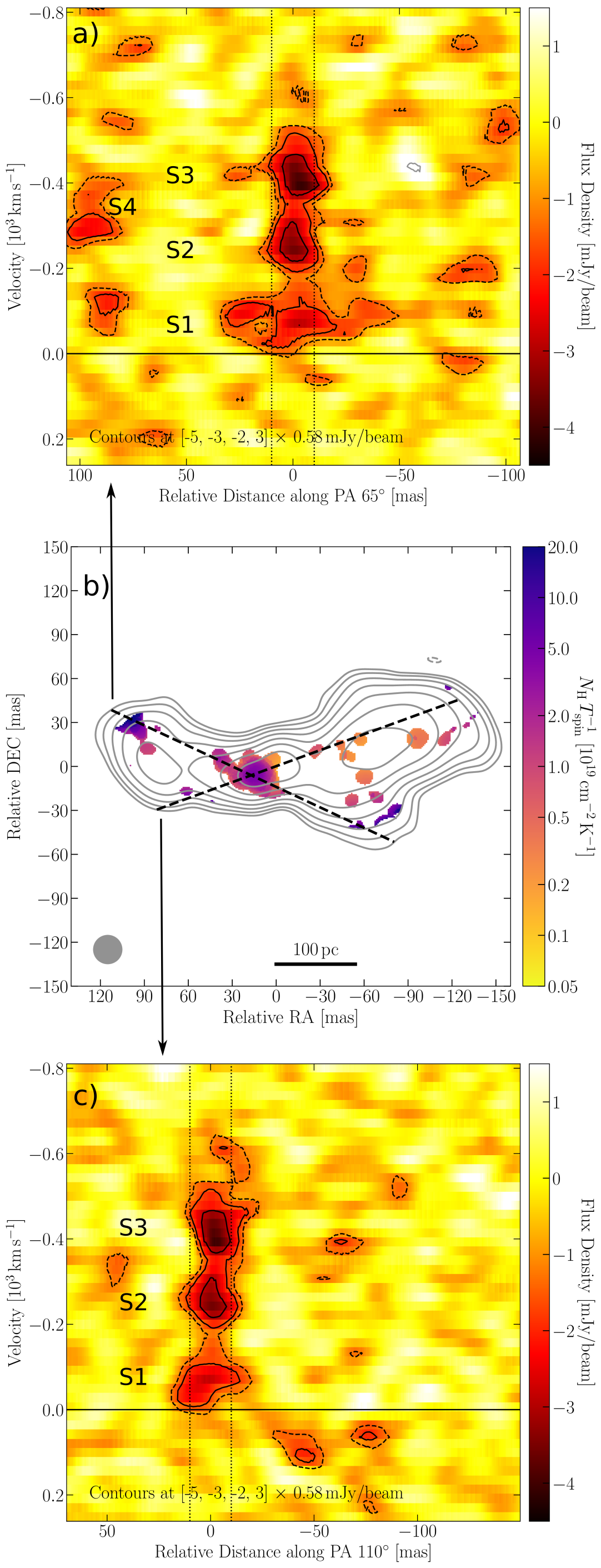}
			\caption{Top panel: position-velocity (PV) diagram of 4C\,52.37 along a cut aligned with the largest extent of the central absorption feature. The solid, horizontal line corresponds to the systemic velocity. The dashed, vertical lines mark the size of the synthesized beam. Contour lines correspond to 3 (grey, solid), -2 (black, dashed), -3 and -5 (black, solid) times the noise level in the cube.  Middle panel: The contour lines show the continuum radio emission as in Fig. \ref{fig:4C52_3C293_cont} with the colormap corresponding to $N_\mathrm{\ion{H}{I}} T^{-1}_\mathrm{spin}$. The dashed lines represent the cuts for the PV diagram. The bottom panel: same as top panel, but a cut along the position angle of W1 and E1. Top and bottom panel are centred on E1.}
			\label{fig:4C52_cont_vr}%
		\end{figure}

\section{Results}
\label{sec:Results}

	\subsection{4C 52.37}
	\label{sec:Results:4C52}
		As we will describe in the sections below, in \object{4C\,52.37} we have succeeded in imaging the radio continuum on scales between $\sim 10$ and 40 pc, 
		and we have recovered a large fraction of the \ion{H}{I} absorption and, in particular, most of the outflowing gas. The outflow appears to be located mostly in the central regions, i.e. in the central $\sim 120$ pc.

		\subsubsection{VLBI Continuum}
		\label{sec:Results:4C52:continuum}

			The top panel of Fig. \ref{fig:4C52_3C293_cont} shows the continuum VLBI image of \object{4C\,52.37} at a resolution of 20\,mas  
			and a zoom-in into the central 200\,pc at 6\,mas. At the higher angular resolution, the central region is clearly resolved into two distinct features E1 and W1. They are located about 20\,mas apart and separated by a small drop in surface brightness. The western feature W1 represents the peak of the radio emission and is compact while the eastern one (E1) is marginally resolved at 6\,mas resolution. Additionally, there are two prominent radio lobes E2 and W2. The overall radio morphology is asymmetric and spans a projected linear size $\sim280\mathrm{\,mas}$ which corresponds to $\sim 530\mathrm{\,pc}$
			
			The majority of the integrated flux density stems from extended radio emission of W2 and E2 with W2 accounting for almost 50\%. The central features E1 and W1 contains about 25\% of the total flux density. In total, we measure a flux density of $429\mathrm{\,mJy}$ which is about 20\% below the value obtained, at lower resolution, by \cite{Gereb2015} and \cite{Maccagni2017} with the WSRT ($577\mathrm{\,mJy}$) and from the VLA survey FIRST at 1.4\,GHz ($575\mathrm{\,mJy}$, \citealt{Helfand2015}). Even considering our conservative flux density uncertainty of 15\%, this suggests that there is a small amount of extended low-surface brightness emission that is either below our sensitivity or resolved out or both. The shortest baseline of the VLBI array which is about $460\mathrm{\,M\lambda}$ implies that extended emission larger than about $550\mathrm{\,mas}$ cannot be recovered.
			
			The EVN observations at 1.659\,GHz by \cite{deVries2009} provide the best comparison on VLBI scales to our observation in both frequency and resolution. Our measured VLBI flux density is consistent, within the uncertainty, with the value obtained by \cite{deVries2009} of $448.3\mathrm{\,mJy}\pm1.1\mathrm{\,mJy}$. Also, the morphology of the source recovered by these two observations matches very well. 

		\subsubsection{The VLBI \ion{H}{I} absorption}
		\label{sec:Results:4C52:spectrum}

			The \ion{H}{I} absorption profile from the archival WSRT observations \citep{Gereb2015} and the integrated spectrum from the new VLBI \ion{H}{I} are shown in Fig. \ref{fig:4C52_spec}. Because of the lower spatial resolution of the WSRT observation, we can use this absorption spectrum as a reference of the total absorption that should have been recovered. The WSRT spectrum comprises of two main features: a narrow, deep absorption close to the systemic velocity and a broad, shallower feature blue-shifted with respect to the systemic velocity. The latter has been interpreted in previous studies as outflowing \ion{H}{I} gas \citep{Gereb2015,Maccagni2017}.
			
			The spatially integrated VLBI \ion{H}{I} absorption spectra have been extracted using the same method as in \cite{Schulz2018}. First, we created a mask based on the continuum image using only those pixels with a value above three times the continuum noise level ($\sigma_\mathrm{VLBI,cont}$). Second, for each channel we applied the mask and integrate either all remaining pixels of the channel (solid, blue line) or integrate only the remaining pixels with a value above three times the noise level of the cube ($\sigma_\mathrm{VLBI,cube}$) leading to the dashed, blue line. 
			The clipped and non-clipped spectra shown in Fig. \ref{fig:4C52_spec} have been obtained by integrating the absorption in the central $60\mathrm{\,mas}\times 40\mathrm{\,mas}$ around E1. This was done to prevent the spectrum be dominated by the noise and because the majority of the absorption is concentrated in this region.
			This would make a comparison between the clipped and non-clipped spectrum meaningless because the latter would have been largely affected by noise. Therefore, we extracted the integrated VLBI spectrum around E1 (see also Sect. \ref{sec:Results:4C52:distribution}).
			
			A comparison between the WSRT and the clipped VLBI spectrum shows (see Fig. \ref{fig:4C52_spec}) that our VLBI observation recovers a large fraction of the \ion{H}{I} broad, blueshifted wing in this region. The narrow feature close to the systemic velocity shows a greater discrepancy between WSRT and VLBI. In the following sections, we argue that the non-detected absorption is largely due to low-column density gas across the source.
			
			As mentioned in \ref{sec:Obs:archival}, we fitted Gaussian distributions using the Python package LMFIT \citep{Newville2014}. We find that both features of the WSRT absorption spectrum are well described by two Gaussian distributions. The fit parameters are listed in Table \ref{tab:Data:fitparam} and plots of the fits are shown in Fig. \ref{fig:4C52_spec_fit}. These parameters are slightly different from \cite{Maccagni2017} because that study used the busy function \citep{Westmeier2014} to fit the spectrum.
			However, the busy function is not suitable to fit the spectra of other sources in our sample because the spectra consist of multiple components. In the WSRT spectrum, the broad feature is blue-shifted by $\sim310\mathrm{\,km\,s^{-1}}$ and it has a full-width at half maximum (FWHM) of about $330\mathrm{\,km\,s^{-1}}$
			
		\subsubsection{\ion{H}{I} gas distribution}
		\label{sec:Results:4C52:distribution}

			Most interestingly, in 4C52.37 we have been able to trace the location of the outflowing \ion{H}{I} gas. The central panel of Fig. \ref{fig:4C52_cont_vr} shows the integrated optical depth of the \ion{H}{I} gas, $\int\tau dv$, detected by our VLBI observation with respect to the continuum source. To calculate $\int\tau dv$, we used the same clipping method as described in the previous section, i.e. a $3 \sigma$ cut in continuum and in the cube. The map shows a large concentration of \ion{H}{I} gas towards the central region, over an elongated region of $\sim 140$ pc, partly overlapping with the eastern component E1, but not W1. In addition, there are smaller clouds of \ion{H}{I} gas distributed over the extent of the continuum emission.
			
			In order to characterize the \ion{H}{I} gas further, we extracted position-velocity diagrams along different position angles. 
			Figure \ref{fig:4C52_cont_vr} shows slices along the extent of the \ion{H}{I} gas in the central region (panel a) and along the position angle of the radio emission in the central region (panel c). 
			The position-velocity plots show that the gas in the central region covers the full range of velocities seen in the spatially integrated absorption spectrum of Fig. \ref{fig:4C52_spec}.
			Three features, labeled S1, S2 and S3, can be seen at increasing (blueshifted) velocity from the systemic velocity.
			All three features are located within the central 100\,pc of the source. While S2 and S3 are unresolved, S1, which corresponds to the gas close to the systemic velocity, is extended (Fig. \ref{fig:4C52_cont_vr}a). 
			From Fig. \ref{fig:4C52_cont_vr} we can therefore derive the location of the outflowing \HI. The gas does not cover the entire central region but it is concentrated towards E1. Since E1 and W1 are of comparable brightness, it would have been possible to detect against the latter a more extended distribution of gas of similar column density if it would be present in front of the continuum.
			
			Figure \ref{fig:4C52_spec} shows that there is a significant fraction of absorption missing close to the systemic velocity in the VLBI extracted around E1, i.e. in the velocity range of S1. The patches of absorption that are depicted in Fig. \ref{fig:4C52_cont_vr}b cannot account for the undetected gas.
			Because they are weak absorption features, we are cautious in interpreting them, but we might be seeing the presence of regular rotating gas is distributed across the radio source. However, we do not reach the optical depth across the radio source to fully recover the gas at the systemic velocity. In comparison,
			most of the absorption in the velocity range of S2 and S3, i.e. the blueshifted component, is recovered in the region around E1. 
			Therefore, clouds of outflowing gas of similar optical depth as S2 and S3 are unlikely to exist anywhere else towards the radio source and the undetected absorption at the systemic velocity must be in the form of compact clouds or diffuse gas of lower optical depth.
			The only possible exception may be the small absorption feature (S4) at the edge of E2 (see Figure \ref{fig:4C52_cont_vr}). This absorption is also blue-shifted with respect to the systemic velocity by about $300\mathrm{\,km\,s^{-1}}$. S4 is much fainter than S2 and S3 and located towards the edge of the radio continuum of E2 rather than the brighter parts. That is why we are cautious in interpreting this feature, but it could be a sign of outflowing clouds outside the central region.

			The peak brightness of the continuum and the noise level of the cube imply an optical depth limit of about 0.035 at $3\sigma_\mathrm{VLBI,cube}$ ($\sim 0.011$ at $1\sigma_\mathrm{VLBI,cube}$). The optical depth limit towards the peak of the western radio lobe (W2) is only slightly higher. W2 has a peak brightness of about $42\mathrm{\,mJy\,beam^{-1}}$ which corresponds to an optical depth limit of 0.042 at $3\sigma_\mathrm{VLBI,cube}$. At the location of E1 the peak optical depth of the absorption feature is about 0.06. Therefore, we would have been able to detect clouds such as S1 at the peak of the western radio lobe (W2) unless the gas clouds have a lower optical depth or the gas is diffuse. This is an indication that the missing \ion{H}{I} absorption is in part due to low-column density gas that is distributed across the continuum source (see Sect. \ref{sec:Results:4C52}).
			
			\begin{figure}
				\centering
				\includegraphics[width=0.97\linewidth]{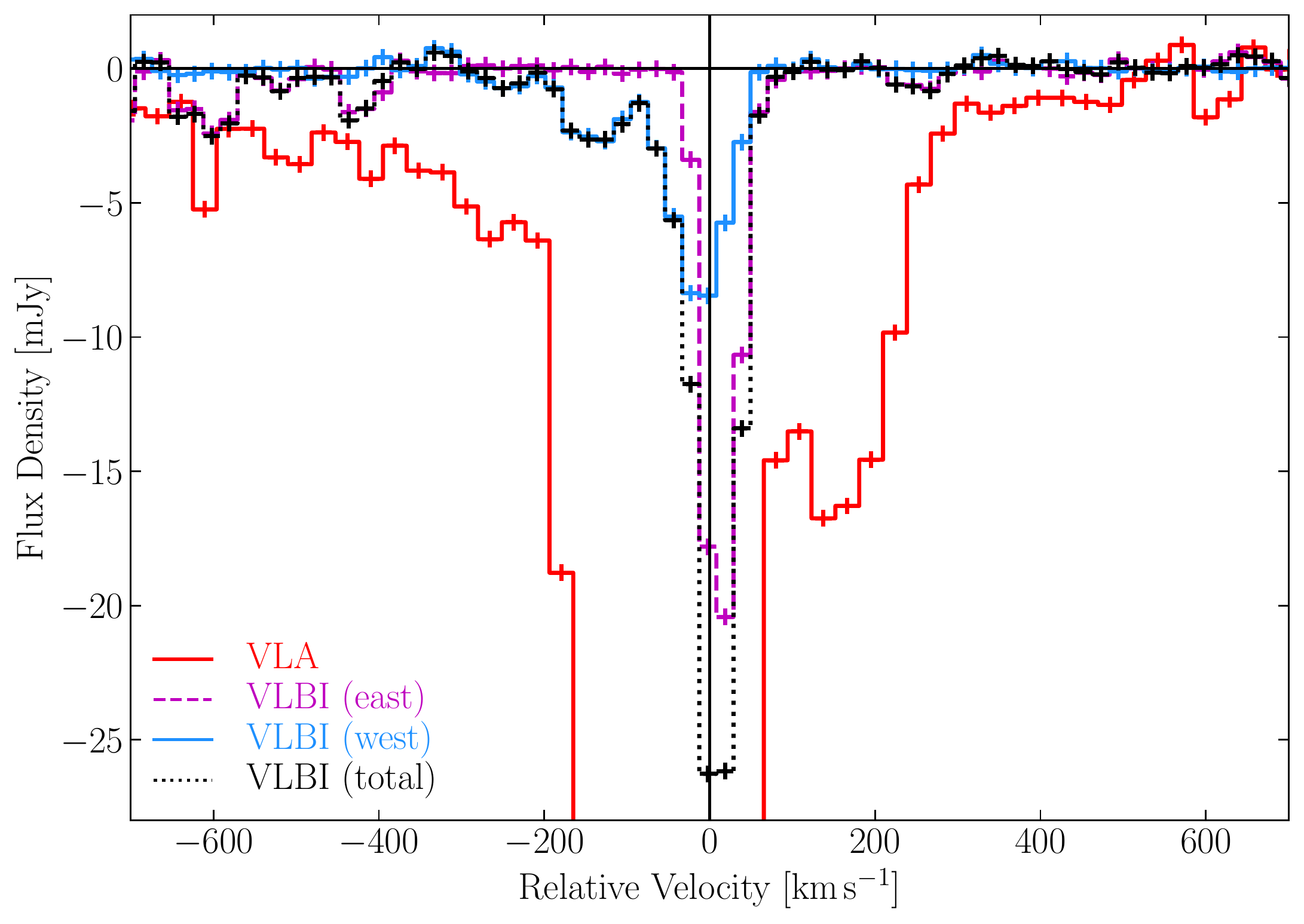}
				\caption{\ion{H}{I} absorption spectra of \object{3C\,293}. The low-resolution VLA spectrum is shown in red. The solid blue and dashed violet lines correspond to the spatially integrated VLBI spectra of the eastern and western part of the source. The dotted, gray line represents the total integrated VLBI spectrum. The VLBI spectra were obtained by integrating only pixels with a value of $\leq-3\sigma_\mathrm{VLBI,cube}$. }
				\label{fig:3C293_spec}%
			\end{figure}

	\subsection{3C\,293}
	\label{sec:Results:3C293}

            Given the much larger spatial scale, we managed to recover only a relatively small part of the continuum structure of \object{3C\,293}. Similarly, also a small fraction of the \ion{H}{I} absorption is recovered. This is particularly the case for the \ion{H}{I} outflow. However, the data show that at least a small fraction of the outflow can be seen against the central ($\sim 50$ pc) region.

		\subsubsection{VLBI continuum}

			The bottom panel of Fig. \ref{fig:4C52_3C293_cont} shows the VLBI continuum emission recovered by our observations. The radio emission from the eastern and western structure spans over 2\arcsec as seen in other high-resolution radio images (e.g., \citealt{Beswick2004,Mahony2013}). The total flux density from our image is about $539\mathrm{\,mJy}$ which is about 17\% of the total flux density recovered by \cite{Beswick2004} at a resolution of $30\mathrm{\,mas}$. The difference can be best explained by the spatial filtering. The shortest baseline of our VLBI array is $400\mathrm{\,M\lambda}$, which means that any extended emission on scales larger than about $630\mathrm{\,mas}$ is resolved out. In contrast, \cite{Beswick2004} combined a global VLBI array similar to ours with MERLIN and VLA data. This has provided significantly shorter baselines which improves the sensitivity for diffuse extended emission. 
			
			The brightest feature in our image matches the location of the VLBI core identified in \cite{Beswick2004}. Interestingly, we find the peak brightness to be a factor 4.1 brighter than in \cite{Beswick2004}. This cannot be explained by the difference in resolution, because the synthesized beam is similar: 25\,mas in our case, compared to 30\,mas in \cite{Beswick2004}. A frequency-dependent effect can also be excluded, because the frequency setup of both studies is chosen to investigate \ion{H}{I} absorption. This suggests that the emission from the radio core is variable and the VLBI core is brighter in our observation compared to previous. Our data does not allow us to test whether also the radio emission extending from the core show changes in brightness or distribution. However, this result warrants caution when comparing observations.

			\begin{figure*}
				\includegraphics[width=0.97\linewidth]{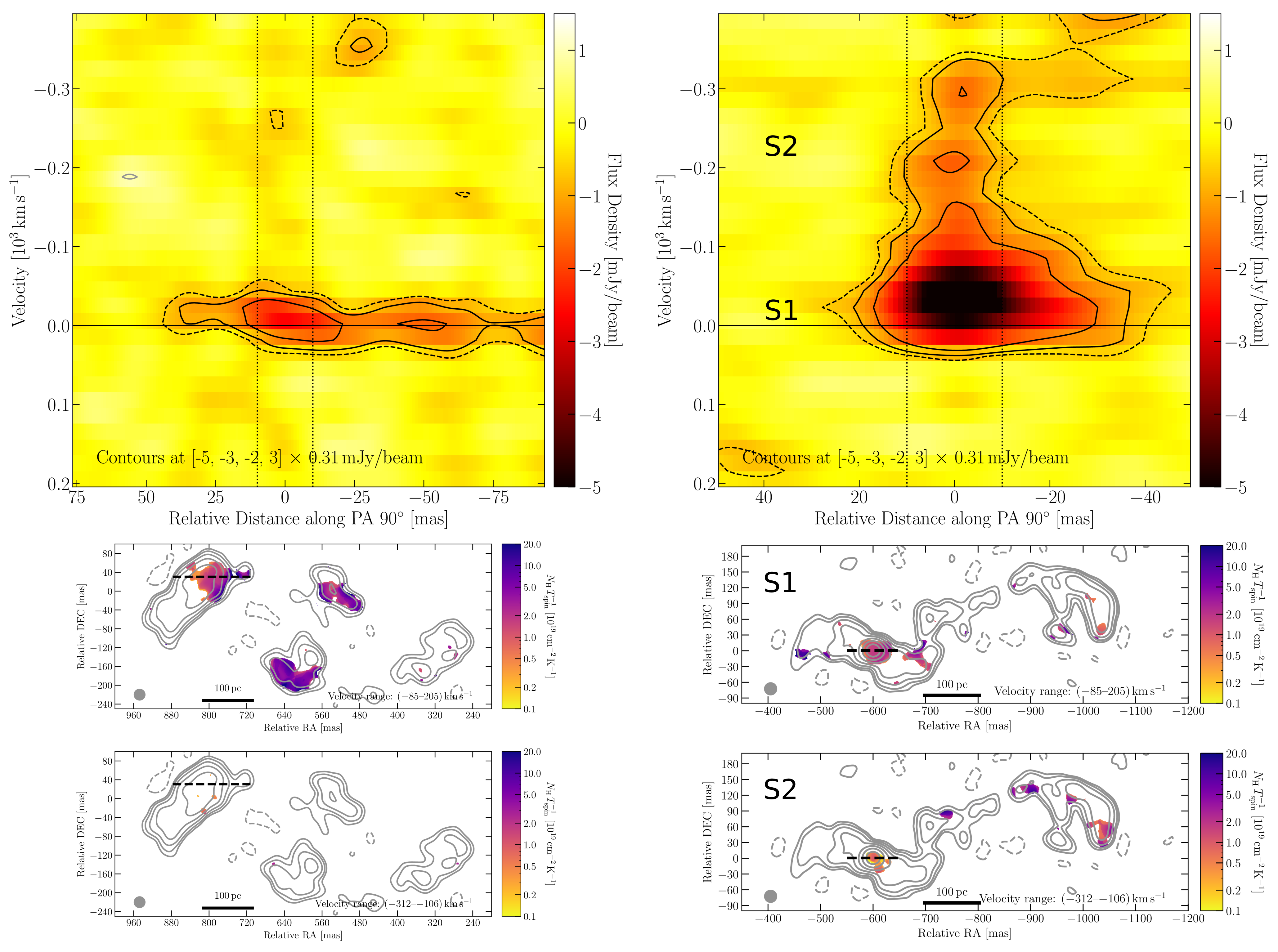}
				\caption{Left panel: position-velocity diagram of a representative location in the eastern part and column density normalized by spin temperature (bottom panel) for the same velocity range as in the right panel. Right panel: position-velocity diagram of \object{3C\,293} at the location of the VLBI core (top panel) and column density normalized by spin temperature separated into systemic and outflowing \ion{H}{I} gas (bottom two panel). The dashed lines in the top two panels correspond to the size of the synthesized beam. The dashed lines in the bottom panels correspond to the location of the slice for the position-velocity diagram.}
				\label{fig:3C293_collage}%
			\end{figure*}

		\subsubsection{The VLBI \ion{H}{I} absorption}

			Figure \ref{fig:3C293_spec} shows the \ion{H}{I} absorption spectrum from the VLBI observations over-plotted to the  spectrum obtained with the VLA with 1\arcsec\ resolution by \cite{Mahony2013}. At this lower resolution, the structure of the source is made of two - barely resolved - components, i.e. eastern and western radio lobes. \cite{Mahony2013} extracted \ion{H}{I} absorption spectra for the eastern and western part, separately. For comparison, we provide spatially integrated \ion{H}{I} absorption spectra from our VLBI data for the east and western part. 
			
			Overall, we recover a relatively small fraction of the VLA-detected absorption. In particular, we recover absorption from the narrow, deep absorption on both lobes. We also marginally detect, on the western side of the source, clouds blue-shifted by up to $-300\mathrm{\,km\,s^{-1}}$ with respect to the systemic velocity (see solid blue line in Fig \ref{fig:3C293_spec}). We do not consider any features below a relative velocity of $-450\mathrm{\,km\,s^{-1}}$. The issue with the frequency setting of our observation mentioned in Sect. \ref{sec:Obs:3C293} strongly affected our ability to perform an optimal continuum subtraction at the corresponding frequencies. As a result, our effectively usable bandwidth is comparable to the one of \cite{Beswick2004}.
			
			The spatial distribution of the \ion{H}{I} absorption obtained by our VLBI observation is shown Fig. \ref{fig:3C293_collage}. On the eastern side, we detect \ion{H}{I} gas distributed in three connected regions. The kinematic of the gas in this part of the source is mostly consistent with quiescent gas like the one originating from the narrow, deep absorption feature. On the western side, the \ion{H}{I} distribution is more patchy. Most of the broad absorption is located towards the core and shows kinematics deviating from the one of the regularly rotating gas (as described in \cite{Beswick2004}). 
			
			We identify two kinematic components (S1 and S2). S1 is likely to be part of the narrow absorption feature similar to the gas recovered on the eastern side and traced also by the observations of \cite{Beswick2004}. S2 is broader in velocity than S1 and blue-shifted by up to $-300\mathrm{\,km\,s}$ with respect to the systemic velocity. Interestingly, S2 appears to be broader than what detected by \cite{Beswick2004} at the location of the core (see their Fig 8). Thus, this suggests that the gas originating the S2 absorption has a highly disturbed kinematics and could be part, as suggested by \cite{Haschick1985}, of the broad \ion{H}{I} outflow observed by \cite{Morganti2005,Mahony2013}. If this is the case, also in \object{3C\,293} the outflow can be observed already in the inner nuclear region ($\leqslant 40$ pc).
			
			We also tentatively detect gas clouds of similar velocity to S2 at other locations in the western part of the source (see Fig. \ref{fig:3C293_collage}). They are at the edges of the recovered radio emission which is why we are cautious with interpreting them. However, if confirmed they would indicate that the outflow is extended as suggested by \cite{Mahony2013} and clumpy.
			
			The gas related to S1 has a peak optical depth of about 0.1 while the gas of S2 peaks at an optical depth of about 0.024. Only at the location of the core, we are able to reach the necessary optical depth of about 0.005 and 0.014 at $1\sigma_\mathrm{VLBI,cube}$ and $3\sigma_\mathrm{VLBI,cube}$, respectively, to detect S2. We cannot detect clouds such as S2 towards the eastern part of the source. The tentatively detected gas cloud at the edge of the western lobe has a peak optical depth of about 0.22. If such clouds exists towards other locations of the western radio lobe where the continuum emission is at the same level or higher, we would have been able to detect them.

\section{Discussion}
\label{sec:Discussion}

	In both radio galaxies, \object{4C\,52.37} and \object{3C\,293}, we were able to  recover (at least partly) and trace the \ion{H}{I} outflows known from previous observations on scales of about 20 pc. We discuss below their structure as revealed on a few tens of pc scale, their properties and their location. We then compare them with already published \ion{H}{I} VLBI results (see Sec. \ref{sec:Discussion:Sample}).

	\subsection{The properties of the \ion{H}{I} gas and the outflow in 4C\,52.37}
	\label{sec:Discussion:4C52HI}
	
		As described in Sec. \ref{sec:Obs:4C52}, in \object{4C\,52.37} we recover most of the \ion{H}{I} outflow and only part of the \ion{H}{I} close to the systemic velocity.  The outflowing gas is  mainly concentrated in the central region (see Fig. \ref{fig:4C52_cont_vr}) and largely contained within two clouds S2 and S3 detected towards E1 (see Fig. \ref{fig:4C52_3C293_cont}), the eastern component of the central region. 
		The location of the core in \object{4C\,52.37} is still unclear and this has implication for understanding the origin of the outflow. This will be further discussed in Sec. \ref{sec:Discussion:4C52Cont}. However, it is reasonable to assume that the active black hole is located either in E1 or W1, which are separated about 40\,mas ($\sim  76$\,pc) in projected size. 
		Table \ref{tab:Data:Outflow} lists the properties of \ion{H}{I} gas in 4C\,52.37 based on our observation and using the same clipping as for the absorption profile. We determine the spin temperature ($T_\mathrm{spin}$) normalized column density ($N_\mathrm{\ion{H}{I}}$) of the outflow by integrating over S2 and S3. In order to calculate any subsequent parameters, we assume $T_\mathrm{spin}=1000\mathrm{\,K}$ which is reasonable considering the proximity of the gas to the AGN. This suggests a column density of $\sim2.0\times 10^{22}\mathrm{cm^{-2}}$. For a spherical geometry, this yields an \ion{H}{I} mass of $1.8\times 10^5\mathrm{\,M_\sun}$. Following \cite{Heckman2002}, we estimate the mass outflow rate to be $\sim4.3\mathrm{\,M_\sun\,yr^{-1}}$ with
		\begin{align}
			\dot{M}_\mathrm{\ion{H}{I}} \sim 30 \frac{r_\star}{\mathrm{kpc}}\frac{N_\mathrm{\ion{H}{I}}}{10^{21}\mathrm{\,cm^{-2}}}\frac{v}{300\mathrm{\,km\,s^{-1}}}\frac{\Omega}{4\pi}M_\sun\mathrm{\,yr^{-1}}
		\end{align}
		where $r_\star$ is the deprojected distance of the cloud, $v$ its velocity and $\Omega$ its solid angle which is assumed to be $\pi$ (see Tab. \ref{tab:Data:Outflow}). We take the synthesis beam as an upper limit on the radius of the clouds because S2 and S3 are both unresolved (see Fig. \ref{fig:4C52_cont_vr}). 
				
	    The gas detected with velocities close to the systemic (S1) represents a fraction of what has been detected by the WSRT. Assuming also for this gas a $T_\mathrm{spin}=1000\mathrm{\,K}$, given the proximity to the AGN, we estimate an average column density of $1.0\times 10^{22}\mathrm{cm^{-2}}$. Based on our VLBI data, we describe S1 by an ellipsoid that is resolved in one direction ($95\,\mathrm{pc}\times\lesssim38\mathrm{\,pc}$) and estimate a lower limit on the mass of the \ion{H}{I} gas of $\sim 2.3\times 10^5\mathrm{\,M_\sun}$. 
		In the kinematics of this gas, we do not detect signatures of rotation. However, we cannot fully rule out the presence of a circumnuclear \ion{H}{I} disk, e.g., given the large fraction of missing absorbed flux. 
		Circumnuclear disks of \ion{H}{I} gas have been traced in the center of a number of radio galaxies such as \object{3C\,236} \citep{Schilizzi2001,Tremblay2010}, Cygnus A \citep{Struve2010}, \object{NGC\,4261} \citep{Jones2000}, Centaurus A \citep{vanGorkom1990,Morganti2008}.
		
		Our observations are limited by the optical depth that we can probe. While we are only able to probe an optical depth similar to E1 towards the peak of the western lobe (W2), S1 is extended and detected against fainter parts around E1. If there is a disk made of clouds like S1, we should have detected it against W1. Therefore, it is possible that the missing absorption comes from extended diffuse component of the systemic gas or lower-column density clouds. The patches of absorption observed across the radio source could be related to this. 
		
		If an \ion{H}{I} disk is present, the absorption recovered in the central region of the radio source would correspond to gas in an inner denser disk, while the undetected absorption would be related to the extended (and lower column density) outer part of the disk. It is possible that the radio jets are drilling into this gas disk. This could explain the asymmetric morphology of the radio emission. Radio jets which are expanding into the ISM have been observed for example in \object{IC\,5063} \cite{Oosterloo2017} and \object{NGC\,1167} \citep{Murthy2019}. In both objects VLBI observation were only able to recover a small fraction of the absorption.
		
	\subsection{Origin of the \ion{H}{I} outflow in 4C\,52.37}
	\label{sec:Discussion:4C52Cont}

		The fact that most of the \ion{H}{I} outflow in \object{4C\,52.37} is concentrated in a very small region in the center of the radio source is important for understanding the driving mechanism of the outflow and the evolution of the radio source. As mentioned in Sect. \ref{sec:Intro}, \object{4C\,52.37} is classified as a low-excitation radio galaxy. As such, the most likely driver of the outflow are the radio jets rather than radiative winds from the AGN. It is worth taking a closer look at the region of E1 and W1 where the AGN is likely to be located.

		We compare the location and spectral index of W1 and E1 (Fig. \ref{fig:4C52_3C293_cont}) using the Gaussian model components fitted to the visibility data at 1.659\,GHz and 4.99\,GHz by \cite{deVries2009}. 
		We find that the position of the model components at 1.659\,GHz and 4.99\,GHz match for W1 and E1. We also fit Gaussian components to W1 and E1 of our VLBI observation and find no difference in location with respect to the components from \cite{deVries2009} at 1.659\,GHz. In addition, we determine the spectral index $\alpha$ between the two frequencies for W1 and E1 to be 0.17 and -0.16, respectively. Both features seem to have a flat spectral index ($|\alpha| < 0.5$). Due to the time span between our observation and the data from \cite{deVries2009}, we do not determine the spectral index between these datasets. 
		As the radio jets of \object{4C\,52.37} seem to be oriented in the plane of the sky and we find \ion{H}{I} absorption towards the region of E1 extending marginally W1, it is possible that the spectral index measurements at least of E1 are affected by an external absorber. This would be similar to the case of for example \object{NGC\,1052} \citep{Kameno2001,Kadler2004b,Sawada-Satoh2008,Baczko2019} where free-free absorption is observed towards the origin of the radio jets.
		An absorber might also explain the small emission gap between E1 and W1 visible in the 6\,mas-resolution image (Fig. \ref{fig:4C52_3C293_cont}). Our VLBI data and \cite{deVries2009} show W1 to be the brightest feature, but this may not be the case intrinsically if free-free absorption would affect our measurements of E1. However, free-free absorption is frequency dependent and as such we should observe a change in position of the brightest feature in the components from \cite{deVries2009} as is observed in \object{NGC\,1052}.
		Simultaneous multi-frequency VLBI observations are necessary to characterize the nature of the central region of \object{4C\,52.37} in greater detail.

		\object{4C\,52.37} has been included in the CORALZ sample \citep{Snellen2004} of candidate young radio sources based on its CSS-like spectrum and is considered to be a young radio source. We can use the same approach as \cite{ODea2001} to estimate a lower limit on its age based on the size of the radio source. As the source extends further to the west, we take the western lobe as a reference estimating its size to be about 130\,mas (250\,pc). This yields a minimum age of the radio source of $t_\mathrm{min}\approx2\times 10^4\mathrm{\,yr}$.
		Thus, the jet would not have been interacting with the ISM over a long period of time to disperse the gas more. 
		This is consistent with the location of the \ion{H}{I} outflow toward E1 close to the center of the AGN.  
		However, the fact that we do not recover all the \ion{H}{I} gas at the systemic velocity, suggests that this gas is distributed over a larger area where it has not been affected by the jet yet. This and the cloud-like nature of the outflow would be consistent with numerical simulations by \cite{Wagner2012,Mukherjee2017,Mukherjee2018}. These simulations have also shown that the propagation of the jet into the ISM and the amount of gas that is affected by it, is determined by gas density, jet power and inclination angle.
		
	\subsection{\ion{H}{I} outflow in 3C\,293}
	\label{sec:Discussion:3C293HI}

		In Sect. \ref{sec:Results:3C293} we have shown that, unlike for \object{4C\,52.37}, our VLBI observation of \object{3C\,293} recovers a relatively small fraction of the \ion{H}{I} absorption and, in particular, of the broad, blueshifted component. We strongly detect signatures of this component against the VLBI core and we tentatively detect clouds against other parts of the western radio lobe. We are cautious with interpreting the latter features, which is why we only derive physical values for the absorption against the VLBI core.
		At this location, we detect unresolved clouds of outflowing \ion{H}{I} gas with velocities blueshifted and deviating up to $300\mathrm{\,km\,s^{-1}}$ with respect to the systemic velocity (S2). The clouds have column densities of $\sim4\times 10^{21}\mathrm{cm^{-2}}$. If we take the beam as an upper limit on the size and assuming a spherical geometry, we derive an \ion{H}{I} mass of the clouds of $1.5\times 10^4\mathrm{\,M_\sun}$ and a mass outflow rate of $\sim0.36\mathrm{\,M_\sun\,yr^{-1}}$.
	
		\cite{Mahony2013} investigate the \ion{H}{I} outflow using high-resolution VLA observations. The study concluded that the outflow is largely extended across the western lobe.  
		The authors state that it is unlikely that all of the outflow is concentrated within the VLBI core because of the implied high optical depth and the location of the \ion{H}{I} outflow in the VLA image. Our findings support and are complementary to these results. The mass outflow rate of S2 is a fraction of the total rate estimated by \cite{Mahony2013} to be $8-50\mathrm{\,M_\sun\,yr^{-1}}$ . We recover only a fraction of the radio emission from the lobe and, at the high spatial resolution of our observations, the core appears to be the brightest component in the western lobe. This suggests we are limited in our sensitivity to trace the outflow in the western lobe. However, our observations show that a fraction of the outflow exist on nuclear scales close to the AGN. Furthermore, we also find clouds at other locations that could be part of the outflow (see bottom right panel of Fig. \ref{fig:3C293_collage}) indicating that the outflowing gas might be extended and clumpy. The optical depth of these tentative detections suggests that we should have been able to find absorption against other parts of the western radio continuum emission, if distributed in clouds with similar properties. The lack thereof suggests that the outflow might have a diffuse component.
		
		The cold ISM gas in \object{3C\,293} is not the only ISM phase that has an outflowing gas component: \cite{Emonts2005} and \cite{Mahony2016} reported outflowing ionized gas. The latter study showed that disturbed ionized gas is observed for a few kpc also in the direction perpendicular to the radio jet.  This has been interpreted as the result of a cocoon of shocked gas producing by the radio jet, expanding perpendicular to the jet propagation into the gas disc aligned with the dust-lane known to be present in the host galaxy of \object{3C\,293}. 
		Such cocoons have been predicted in numerical simulations describing the interaction between a newly formed radio jet and the surrounding, clumpy medium (e.g.,\citealt{Wagner2011,Mukherjee2016,Mukherjee2017}). 
				
		Our results are consistent with this scenario which predicts that the interaction between jet and ISM begins as soon as the jet starts expanding into the gas and it occurs along and perpendicular to the jet propagation. Additionally, the longer the jet expands into the ISM the more the gas can get dispersed. The detection of clouds of outflowing \ion{H}{I} towards the core in addition to outflowing gas distributed across the western lobe further support this scenario. 
		
	\subsection{Properties of \ion{H}{I} outflows in our sample}
	\label{sec:Discussion:Sample}

		We now compare the results on the properties of the \ion{H}{I} outflows presented here for \object{4C\,52.37} and \object{3C\,293} with the other two objects (\object{4C\,12.50} and \object{3C\,236}) presented in \cite{Morganti2013} and \cite{Schulz2018}. These four radio galaxies are the only objects for which fast outflows in \ion{H}{I} absorption on milliarcsecond scales have been studied.
				
		We can summarise the main results we have obtained as:
		\begin{enumerate}
			\item We recover at least part of the \ion{H}{I} outflow on VLBI (20 pc) scale in all objects. 
			\item This indicates that all the outflows include at least a component of relatively compact clouds (with masses in the range of $10^4-10^5M_\sun$)
			\item We suggest that the clouds might be embedded in a more diffuse component. We detect this in \object{4C\,12.50} \citep{Morganti2013}. Also in \object{3C\,236} we concluded that at least part of the undetected the outflow is made of a diffuse, lower column density component \citep{Schulz2018}. This could also be the case in \object{3C\,293} since the outflow must be extended over the western lobe.
			\item The data alone provide only limited additional insight into the understanding of the driving mechanism. As already mentioned in the introduction, we remark that the radio galaxies  are low luminosity optical AGN (with the exception of 4C~12.50), and are all young or restarted radio galaxies. Therefore, the radio jet is a very likely mechanism for driving the outflows. The properties of the outflows we derive are consistent with what is predicted by the simulation of jet-driven outflow.  The derived density of the clouds are also similar to what used in the simulations (e.g., \citealt{Wagner2011,Wagner2012,Mukherjee2016,Mukherjee2018b}).
		\end{enumerate}
			
		In all four sources we find evidence for a clumpy distribution of the outflowing \ion{H}{I} but at different levels. This is shown in the spatial and velocity distribution of the gas as well as a the difference between the blueshifted \ion{H}{I} absorption recovered at low and high angular resolution.  
		Here, we attempt to explore whether our results show an evolutionary sequence as expected from numerical simulations by considering the age of the sources and the characteristics of the \ion{H}{I} outflow both in the integrated spectrum as well as in the spatially resolved distribution of the outflowing gas. 
		Numerical simulations (see e.g. \citealt{Wagner2011,Wagner2012}) have shown that radio jets will accelerate and disperse gas along its direction of propagation, but also transversal to it when they expand into a clumpy medium. With time, more and more of the ISM gas will be dispersed  \citep{Mukherjee2018b} leading to changes in the properties of the ISM and outflowing gas. 
		
		In Sect. \ref{sec:Discussion:4C52Cont}, we have estimated a minimum age of the radio source of \object{4C\,52.37} of $t_\mathrm{min}\approx2\times 10^4\mathrm{\,yr}$ based on the size of the radio continuum emission. This is of the same order as the minimum age estimated for \object{4C\,12.50} \citep{Morganti2013}. In \object{4C\,12.50}, all of the \ion{H}{I} outflow has been recovered with VLBI in the form of a slightly extended gas cloud towards the southern jet, while in \object{4C\,52.37} most of the outflow is recovered and located towards the inner region of the radio source. It is still mostly compact in \object{4C\,52.37}, but the undetected absorption in this source suggests that a fraction of the outflowing gas is located elsewhere. Based on this, \object{4C\,12.50} would be consistent with an early evolutionary stage of the jet-ISM interaction and \object{4C\,52.37} would be in slightly more advanced stage compared to \object{4C\,12.50}.
		
		The minimum age of \object{4C\,12.50} and \object{4C\,52.37} is an order of magnitude lower than the age estimated for \object{3C\,236} \citep{ODea2001,Tremblay2010}. In \object{3C\,236}, only a fraction of the \ion{H}{I} outflow was recovered, mostly towards the nuclear region of the radio source, but also towards the hot spot of the south-east jet in the form of an extended cloud. This would be consistent with a later stage in the evolution of the radio source where the jet had significant time to disperse most of the gas. In agreement with this, also in the case of \object{3C\,293} we recover only a fraction of the outflow, mostly at the location of the core.
		
		In order to characterize \ion{H}{I} spectra in more detail, we fitted between 1 and 3 Gaussian distributions to the low and high resolution spectra. The spectra and fit parameters are provided in Appendix \ref{sec:appendix} and Table \ref{tab:Data:fitparam}, respectively. While Gaussian functions do not represent a physical model, they provide a simple and consistent approach. 
		The fits to the VLA spectra of 3C\,236 and 3C\,293 are consistent with the fits presented in \cite{Labiano2013} and \cite{Mahony2013}. In case of 4C\,12.50 we only performed a fit of the WSRT spectrum, because the VLBI spectrum is consistent with it as shown in \cite{Morganti2013}. For 4C\,52.37 our fit parametrization is different from \cite{Maccagni2017}, because this study used the busy function \cite{Westmeier2014} to fit the spectrum.

		The shape of the blue wing in the absorption spectra differ greatly among the four sources. In the case of \object{4C\,12.50} and \object{4C\,52.37} it is more detached from the systemic gas than in \object{3C\,236} and \object{3C\,293}. In the latter two sources, the blue wing is shallow and smooth including a smooth transition from the systemic \ion{H}{I}. The difference in the blue wing is reflected by the property of the Gaussian components that describe that part of the spectrum. The FWHM of \object{3C\,236} (VLA component 3) is larger by factor of 2 than \object{4C\,12.50} (WSRT component 2) and \object{4C\.52.37} (WSRT component 3). The component describing the wing of \object{3C\,293} (VLA component 2) has a FWHM similar to that of for the latter two sources, but the uncertainty of the fit result is very high most likely due to correlation with the parameters of component 3. It is therefore possible that the shape of the blue wing in the \ion{H}{I} spectrum is a signature of the compactness of the \ion{H}{I} outflow.
		
		It is clear that the \text{actual} situation is more complicated, because the above arguments do not consider individual difference in the gas distribution and evolution of the system, e.g., merger activity (\object{3C\,293}) and previous cycles of activities (\object{4C\,12.50}, \object{3C\,236}, and \object{3C\,293}). The ages of the radio sources provide only limited information as these are lower limits. Also, \object{4C\,12.50} is the only source in our sample where no absorption is detected towards the VLBI core. Interestingly, the density of the detected outflowing gas clouds is relatively similar among the four sources and the VLBI observations probe similar spatial scales in all of them. Therefore, it is intriguing that a connection can be drawn in the form of an evolutionary scheme for these sources and it shows the need for further studies of \ion{H}{I} outflows in other systems in combination with numerical simulations.

	\begin{table*}[htpb]
		\centering
		\caption[]{Properties of the kinematically disturbed \ion{H}{I} gas}
		\label{tab:Data:Outflow}
		\begin{tabular}{cccccccccc}
				\hline
				Source\tablefootmark{a} & Component\tablefootmark{b} & $N_\mathrm{\ion{H}{I}}T_\mathrm{spin}^{-1}$\tablefootmark{c} & $N_\mathrm{\ion{H}{I}}$\tablefootmark{d} & $d$\tablefootmark{e} & $n_\mathrm{\ion{H}{I}}$\tablefootmark{f} & $m_\mathrm{\ion{H}{I}}$\tablefootmark{g} & $v$\tablefootmark{h} & $r_\star$\tablefootmark{i} & $\dot{M}_\mathrm{\ion{H}{I}}$\tablefootmark{j}\\
				& & [$10^{19}$\,cm$^{-2}$\,K$^{-1}$] & [$10^{19}$\,cm$^{-2}$] & [pc] &  [cm$^{-3}$] &[$10^4$\,M$_\sun$] & [km\,s$^{-1}$] & [pc] & [M$_\sun$\,yr$^{-1}$] \\
				\hline
				\hline
				\object{4C\,12.50} & Outflow & 4.6 & 460 & $50\times12$ & 150-300 & 1.6 & 1000 & - & 16-29 \\
				\object{4C\,52.37} 	& Outflow  & 2.0 & 2000 & $\lesssim38$ & 170  & 18 & 450 & $\lesssim38$ & 4.3\\
									& Systemic & 1.0 & 1000 & $95\times\lesssim38$ & 85 & 23 & - & - & -\\
				\object{3C\,236} & Outflow (Core) & 0.78 & 780 & $\lesssim36$ & $120$ & 2.8 & 640 & $\lesssim40$ & $0.5$ \\
				\object{3C\,293} & Outflow (Core) & 0.4 & 400 & $\lesssim24$ & 54 & 1.5 & 300 & $\lesssim24$ & 0.36\\
								 & Systemic (Core) & 2.0 & 2000 & $33\times 33$ & 196 & 1.4 & - & - & - \\
				\hline
		\end{tabular}
		\tablefoot{
				\tablefoottext{a}{Values for 4C\,12.50 from \cite{Morganti2013}, for 3C\,236 from \cite{Schulz2018}, for 4C\,52.37 and 3C\,293 from this paper. For 4C\,12.50 and 3C\,236, we focus on the kinematically disturbed \ion{H}{I} gas.} 
				\tablefoottext{b}{For 4C\,52.37, the values for `Outflow' after integrating over S2 and S3.}
				\tablefoottext{c}{Mean \ion{H}{I} column density normalised by spin temperature.}
				\tablefoottext{d}{Mean \ion{H}{I} column density. $T_\mathrm{spin}=1000\mathrm{\,K}$ is assumed except for \object{4C\,12.50} for which \cite{Morganti2013} assumed $T_\mathrm{spin}=100\mathrm{\,K}$.}
				\tablefoottext{e}{Projected size of the components. For the outflow, a spherical geometry is assumed with an upper limit of the diameter based on the synthesized beam. For S1, an ellipsoidal geometry is assumed, with the major and minor axis given.}
				\tablefoottext{f}{Density of the \ion{H}{I} clouds. The same $T_\mathrm{spin}$ values as for the column density are assumed here.}
				\tablefoottext{g}{Mass of the \ion{H}{I} clouds for the chosen $T_\mathrm{spin}$ values.}
				\tablefoottext{h}{Peak velocity of the \ion{H}{I} clouds relative to the peak velocity of S1.}
				\tablefoottext{i}{Deprojected distance of the \ion{H}{I} clouds}
				\tablefoottext{j}{Mass outflow rate following \cite{Heckman2002} for the chosen $T_\mathrm{spin}$ values.}
		}
		\end{table*}
	
\section{Summary \& Conclusion}
\label{sec:Summary}
	
	In this paper, we have presented results, using global spectral-line VLBI, for two additional sources from our investigation of the parsec-scale \ion{H}{I} outflow in four radio galaxies. In the case of 4C\,52.37, this is the first such VLBI study of the \ion{H}{I} gas distribution. We recovered the majority of the \ion{H}{I} absorption related to the outflowing gas and a significant fraction of the systemic \ion{H}{I} gas with a resolution of 20\,mas. We detect the outflow in the form of two unresolved clouds towards the central region of the radio source. They have a combined mass outflow rate of less than $4.3\mathrm{\,M_\sun\,yr^{-1}}$ and cover about $600\mathrm{\,km\,s^{-1}}$. The recovered systemic gas is extended and largely concentrated towards the central 100\,pc, but it shows also signs of being clumpy and distributed across the radio source. We find signs that the radio jet of \object{4C\,52.37} expands into the gas disk, but further observations are necessary.
	
	In case of \object{3C\,293}, our VLBI observation resolved out most of the extended radio emission and it suffered from technical difficulties limiting the frequency range over which we can probe the broad \ion{H}{I} outflow. Nevertheless, we recover a fraction of the \ion{H}{I} gas over a velocity range of $300\mathrm{\,km\,s^{-1}}$ in the form of distinct gas clumps. Some of the clumps are located towards the VLBI core, i.e, close to the AGN, and are likely signs of the outflow in this source. As both, \object{4C\,52.37} and \object{3C\,293} are low-excitation galaxies, it is reasonable to assume that only the radio jets are capable of driving the \ion{H}{I} outflows.
	
	All four sources (\object{4C\,12.50}, \object{4C\,52.37}, \object{3C\,236}, \object{3C\,293}) in our sample have in common that that the outflowing \ion{H}{I} gas is clumpy and that some of the gas is located towards the innermost region {($< 50$ pc)} of the radio source. For \object{4C\,12.50} and \object{4C\,52.37} we find that almost all of the outflowing \ion{H}{I} gas is concentrated in a small number of compact clouds, whereas for the other two sources our data implies that the \ion{H}{I} outflow is more extended and might have a diffuse component. Taking into account our observational limitations, we can interpret the properties of the \ion{H}{I} outflow as a sign of different stages of jet-ISM interaction of these sources. This would be consistent with numerical simulations that have shown that the radio jets would disperse more and more of the ISM gas while they are expanding over time.

	Our results highlight the importance of high-angular resolution observations to understand the gas properties and dynamics in connection to AGN activity. 
	Upcoming facilities, such as the Square Kilometre Array and its pathfinders such as Apertif, MeerKAT and ASKAP (\citealt{Braun2015,Morganti2015b,Oosterloo2010,Gupta2017,Johnston2007}) and the DSA-2000 \citep{Hallinan2019}, will produce blind large-scale surveys of \ion{H}{I} absorption which will increase the number of radio AGN with \ion{H}{I} absorption, particularly at low radio powers. In addition to large-scale \ion{H}{I} surveys to be conducted over the next decade, follow-up VLBI observations with new telescopes such as the next-generation Very Large Array \citep{Murphy2018} and the SKA \citep{Paragi2015}, will be crucial for spatially-resolving \ion{H}{I} absorption against radio jets in order to quantify the impact of jet-ISM feedback on host galaxy properties \citep{Nyland2018a}.  The results from future \ion{H}{I} absorption studies of much larger samples will ultimately inform numerical simulations of jet-ISM interactions (e.g., \citealt{Wagner2012,Mukherjee2016,Mukherjee2018,Mukherjee2018b,Bicknell2018}), thereby providing new insights into the role of jet-driven feedback in the broader context of galaxy evolution.
   
\begin{acknowledgements}
	We thank the referee, Bjorn Emonts, for the insightful comments which helped to improve the manuscript.
	RS gratefully acknowledge support from the European Research Council under the European Union's Seventh Framework Programme (FP/2007-2013)/ERC Advanced Grant RADIOLIFE-320745.
	Basic research in radio astronomy at the U.S. Naval Research Laboratory is supported by 6.1 Base Funding.
	The European VLBI Network is a joint facility of independent European, African, Asian, and North American radio astronomy institutes. Scientific results from data presented in this publication are derived from the following EVN project code: GN002. 
	The National Radio Astronomy Observatory is a facility of the National Science Foundation operated under cooperative agreement by Associated Universities, Inc. The Long Baseline Observatory is a facility of the National Science Foundation operated under cooperative agreement by Associated Universities, Inc.
	The Arecibo Observatory is a facility of the National Science Foundation (NSF) operated by SRI International in alliance with the Universities Space Research Association (USRA) and UMET under a cooperative agreement. The Arecibo Observatory Planetary Radar Program is funded through the National Aeronautics and Space Administration (NASA) Near-Earth Objects Observations program.
	Based on observations made with the NASA/ESA Hubble Space Telescope, and obtained from the Hubble Legacy Archive, which is a collaboration between the Space Telescope Science Institute (STScI/NASA), the Space Telescope European Coordinating Facility (ST-ECF/ESA) and the Canadian Astronomy Data Centre (CADC/NRC/CSA).
	This research has made use of NASA's Astrophysics Data System Bibliographic Services.
	This research has made use of the NASA/IPAC Extragalactic Database (NED) which is operated by the Jet Propulsion Laboratory, California Institute of Technology, under contract with the National Aeronautics and Space Administration.
	This research made use of Astropy, a community-developed core Python package for Astronomy \citep{Astropy2013}.
	This research made use of APLpy, an open-source plotting package for Python \cite{Aplpy2012}.
\end{acknowledgements}

\bibliographystyle{aa} 
\bibliography{References} 


\begin{appendix} 

	\section{Fits to \ion{H}{I} absorption sepctra}
	\label{sec:appendix}

	\begin{table*}[h!]
		\centering
		\caption[]{Parameters of the Gaussian fit to the \ion{H}{I} absorption spectra}
		\label{tab:Data:fitparam}
		\begin{tabular}{cccccccccc}
				\hline
				Source\tablefootmark{a} & Instr.\tablefootmark{b} & Comp.\tablefootmark{c} &$N_\mathrm{free}$\tablefootmark{d} & $\chi^2$\tablefootmark{e} & $\chi^2_\mathrm{red}$\tablefootmark{f} & Center\tablefootmark{g} & FWHM\tablefootmark{h} & Peak\tablefootmark{i} \\
				& & & & & & [km\,s$^{-1}$] & [km\,s$^{-1}$] & [mJy] \\
				\hline
				\hline
				\object{4C\,12.50} 	& WSRT  	 & 1 & 357 & 723.235  & 2.03 & $84.0\pm0.8$ & $135\pm3$ & $41.9\pm0.6$ \\
									&			 & 2 &     &		 &	    & $110\pm0.010$   & $850\pm40$   & $7.0\pm0.4$ \\
									&			 & 3 &     &		 &	    & $-880\pm10$   & $340\pm20$   & $6.4\pm0.4$ \\
				\object{4C\,52.37} 	& WSRT  	 & 1 & 179 & 61.849  & 0.35 & $1\pm3$ & $ 136\pm8$ & $7.0\pm0.3$ \\
									&			 & 2 &     &		 &	    & $-309\pm9$   & $330\pm30$   & $3.9\pm0.2$ \\
									& VLBI  	 & 1 & 105 & 3.021   & 0.03 & $-48\pm1$ & $64\pm3$ & $3.1\pm0.2$ \\
									&			 & 2 &     &		 &	    & $-231\pm2$ & $76\pm6$   & $2.1\pm0.1$ \\
									&			 & 3 &     &		 &	    & $-396\pm2$ & $100\pm5$   & $2.7\pm0.1$ \\
				\object{3C\,236} 	& VLA   	 & 1 & 148 & 943.735 & 6.38 & $66\pm1$ & $78\pm4$ & $67\pm3$    \\
									&			 & 2 &     &		 &	    & $52\pm4$ & $270\pm20$   & $42\pm3$    \\
									&			 & 3 &     &		 &	    & $-210\pm80$   & $700\pm100$     & $6\pm1$     \\
									& VLBI (Core)& 1 & 134 & 25.049	 & 0.19 & $-103.0\pm0.5$ & $60\pm1$ & $2.91\pm0.05$ \\
									& 			 & 2 &     &  	     &      & $-327\pm1$ & $86\pm2$ & $1.89\pm0.04$ \\
									& 			 & 3 &     &  	     &      & $-544\pm2$ & $62\pm4$ & $0.92\pm0.05$ \\
									& VLBI (Lobe)& 1 & 134 & 0.528	 & 0.004 & $61\pm1$ & $88\pm3$ & $54\pm6$    \\
									& 			 & 2 &     &  	     &       & $130\pm70$ & $130\pm70$ & $4\pm3$ \\
									& 			 & 3 &     &  	     &       & $-99\pm3$   & $38\pm6$   & $2.8\pm0.4$     \\
				\object{3C\,293} 	& VLA (west) & 1 & 110 & 776.292 & 7.06 & $-55\pm2$ & $165\pm5$ & $79\pm2$    \\
									&			 & 2 &     &		 &	    & $-400\pm70$   & $400\pm200$     & $4\pm1$     \\
									&			 & 3 &     &		 &	    & $210\pm20$   & $120\pm40$   & $6\pm2$     \\
				\hline
		\end{tabular}
		\tablefoot{
			\tablefoottext{a}{Name of the source}
			\tablefoottext{b}{Instrument used to obtain the spectrum}
			\tablefoottext{c}{Components used to fit the spectrum}
			\tablefoottext{d}{Number of degrees of freedom for the fit}
			\tablefoottext{e}{Chi-sqaured of the fit}
			\tablefoottext{f}{Reduced chi-sqaured of the fit}
			\tablefoottext{g}{Velocity of the center of the Gaussian component}
			\tablefoottext{h}{Full-width at half maximum of the Gaussian component}
			\tablefoottext{i}{Peak of the Gaussian component}
		}
	\end{table*}

		\begin{figure}[h]
			\centering
			\includegraphics[width=0.97\linewidth]{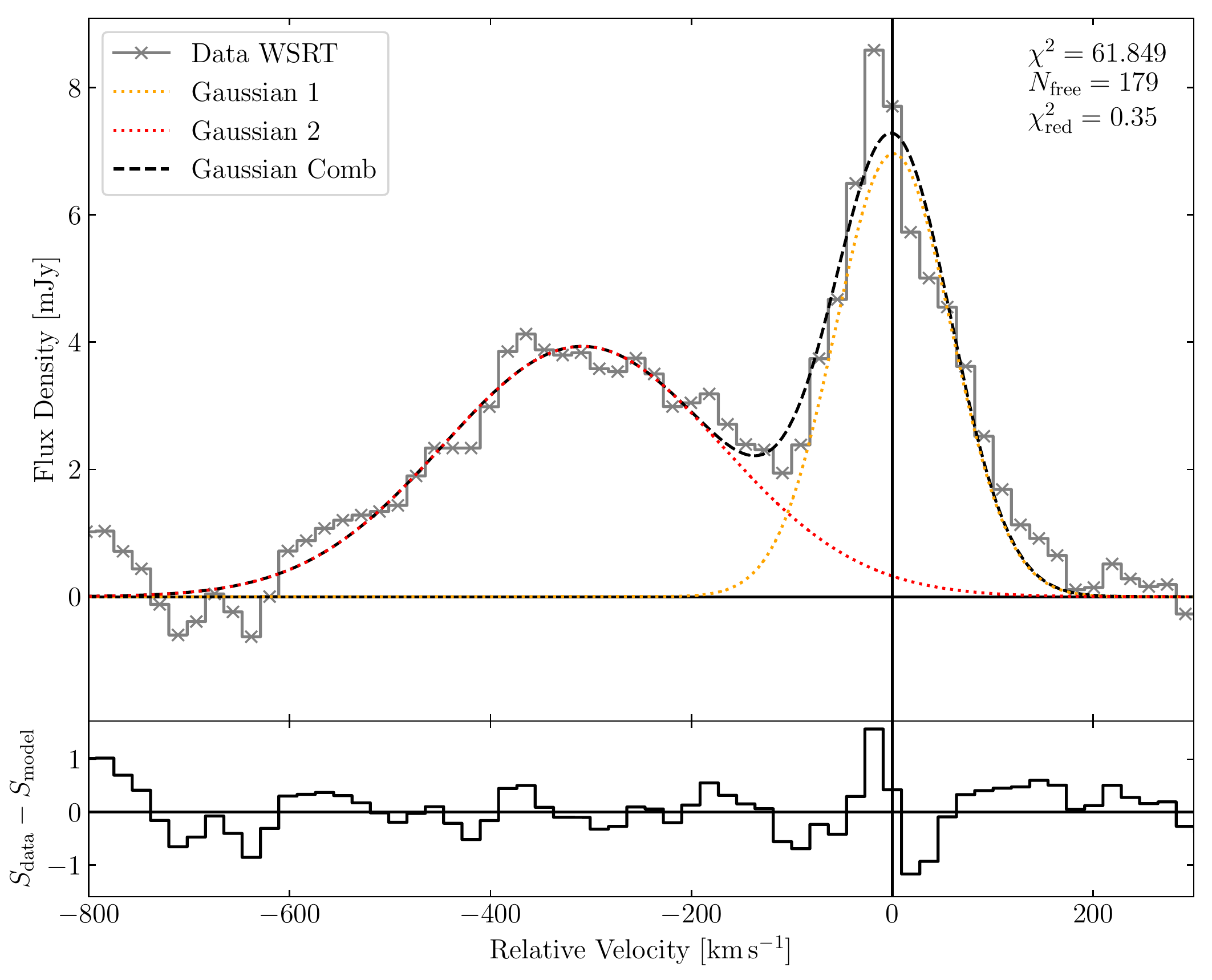}
			\includegraphics[width=0.97\linewidth]{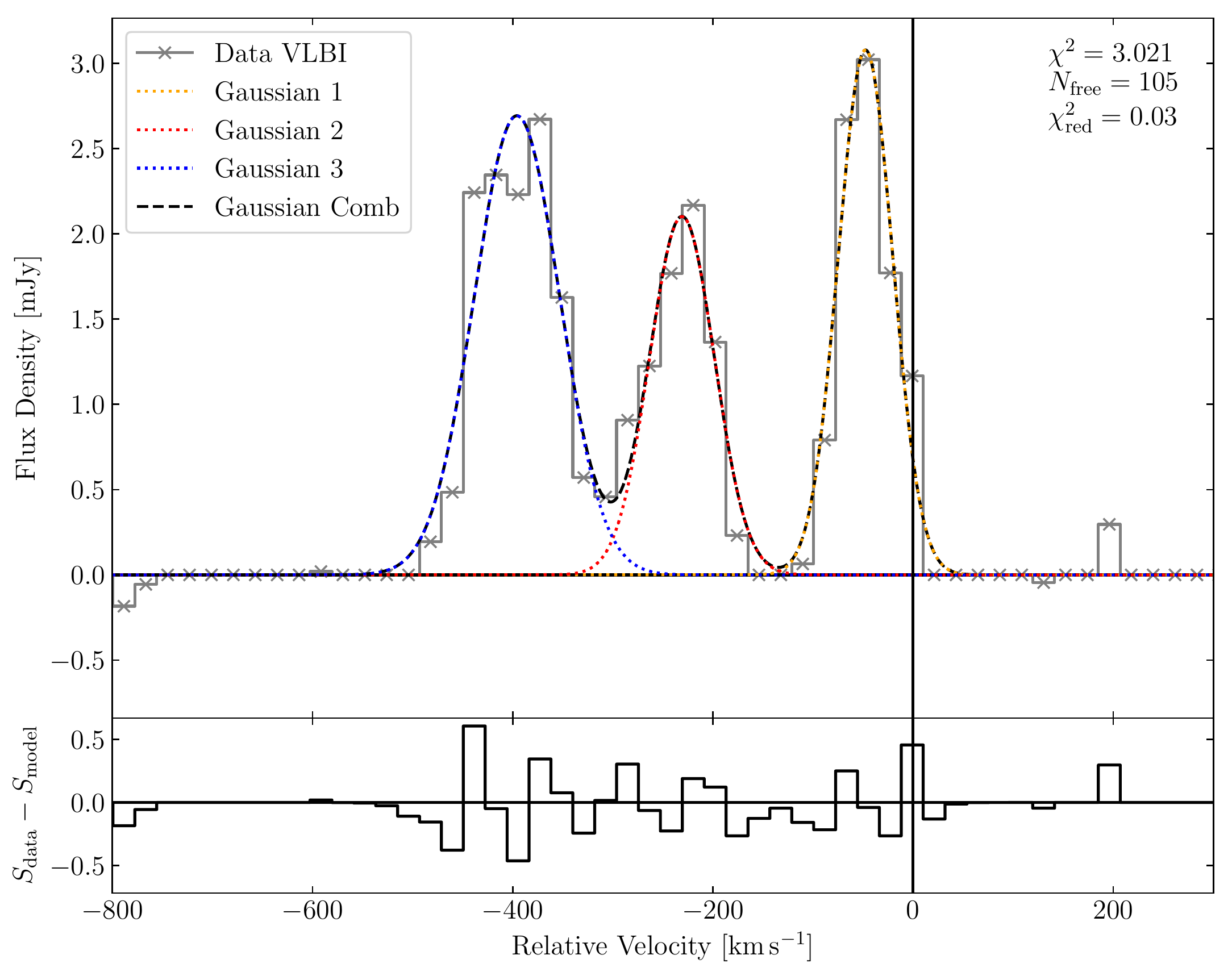}
			\caption{Fits of Gaussian functions to inverted \ion{H}{I} absorption spectra of 4C52.37. The fit paramaters are given in Table \ref{tab:Data:fitparam}. Top: WSRT spectum from Fig. \ref{fig:4C52_spec} with two Gaussian functions fitted to it (red and yellow dotted lines). The combination of the two fit functions is shown as the black dashed line. The data is colored in grey. Bottom: clipped integrated VLBI spectrum from Fig. \ref{fig:4C52_spec}. Three Gaussian components were fitted to it (colored dotted lines) and their combination is shown as the black dashed line.}
			\label{fig:4C52_spec_fit}%
		\end{figure}
		
		\begin{figure}[h]
			\centering
			\includegraphics[width=0.97\linewidth]{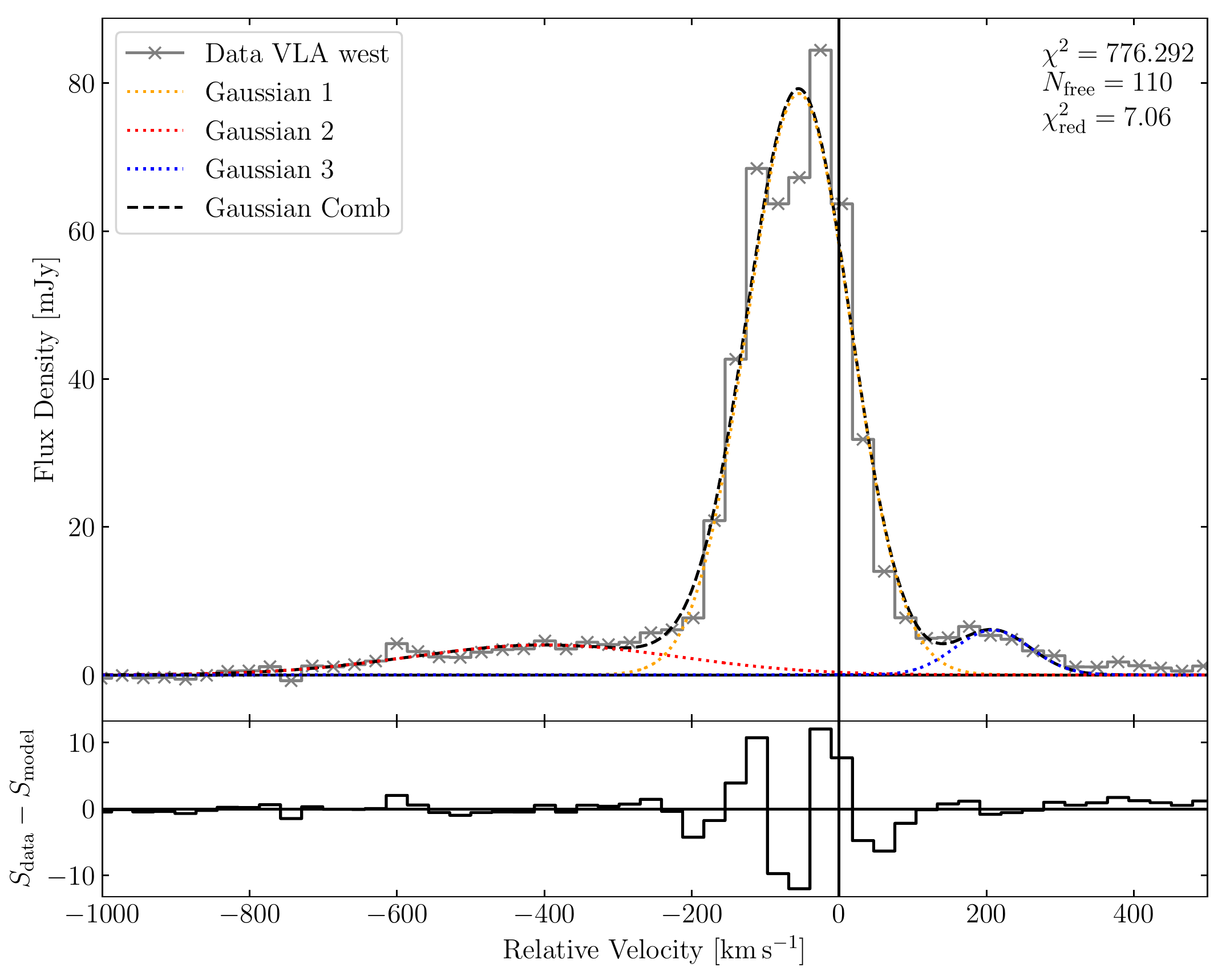}
			\caption{Inverted VLA \ion{H}{I} absorption spectrum of \object{3C\,293} of the western component from Fig. \ref{fig:3C293_spec}. Data from \cite{Mahony2013}. Three Gaussian components were fitted to it (colored dotted lines) and their combination is shown as the black dashed line. Fit parameters are listed in Table \ref{tab:Data:fitparam} }
			\label{fig:3C293_spec_fit}%
		\end{figure}

		\begin{figure}[h]
			\centering
			\includegraphics[width=0.97\linewidth]{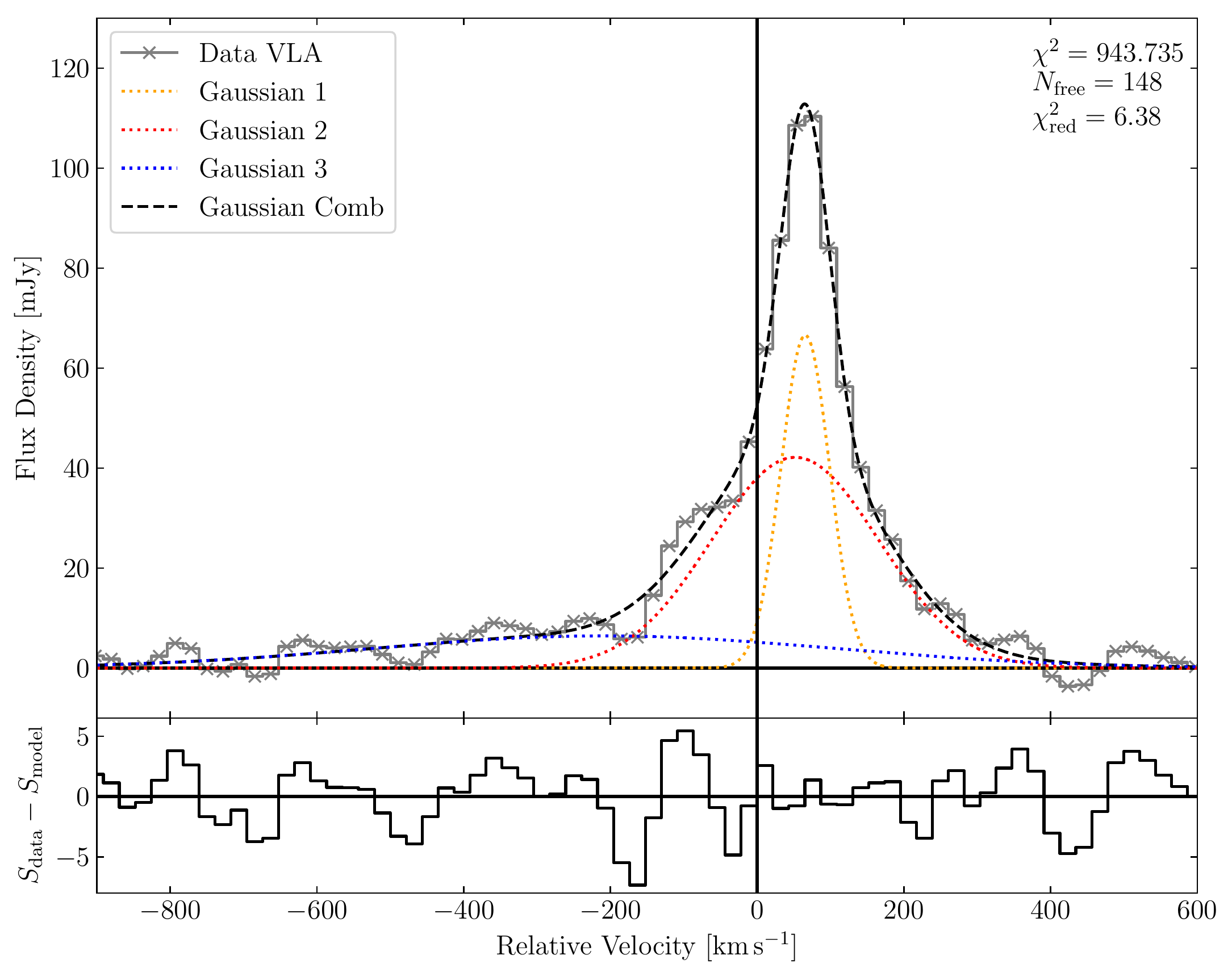}
			\includegraphics[width=0.97\linewidth]{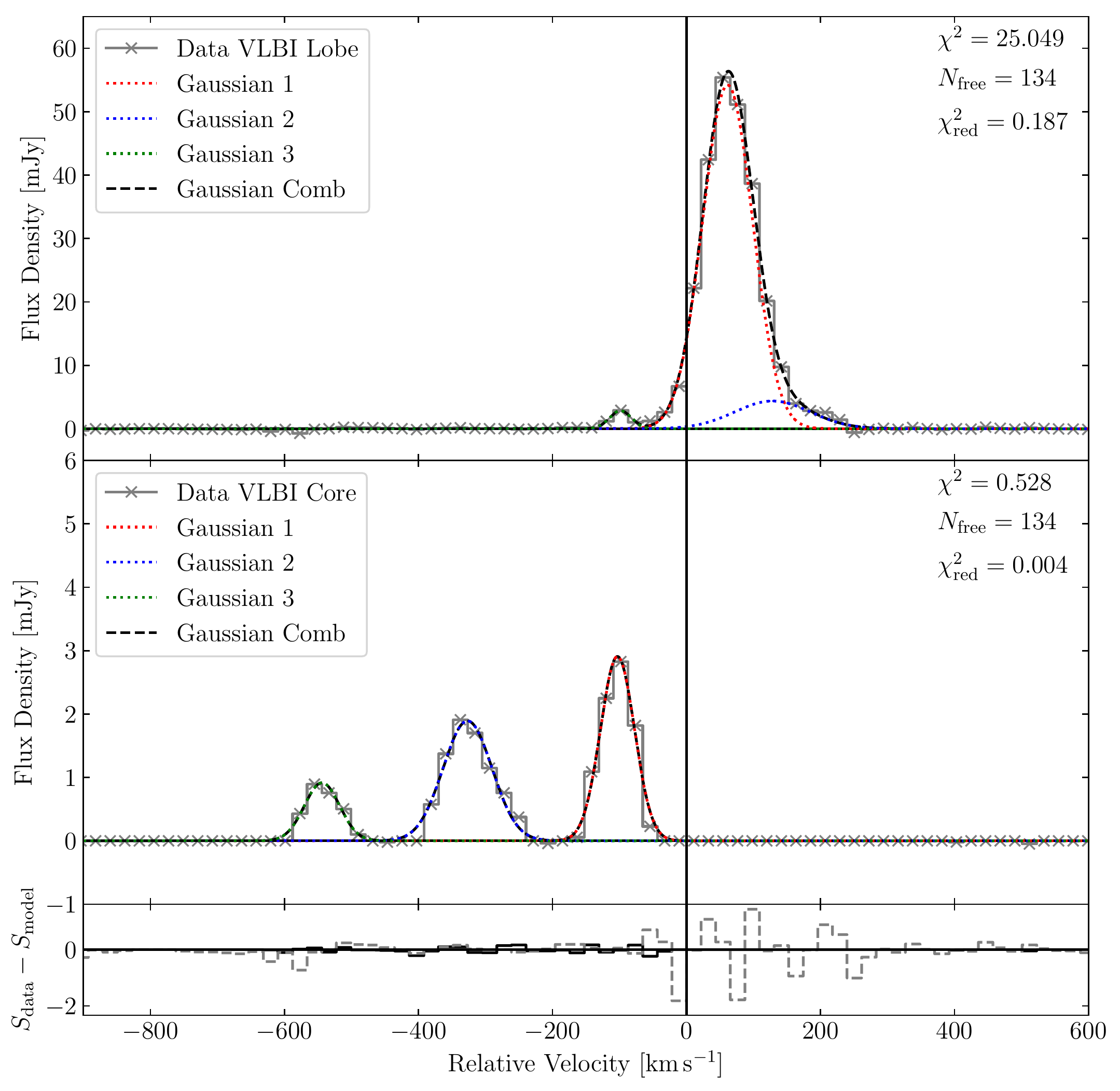}
			\caption{Fits of Gaussian functions to inverted \ion{H}{I} absorption spectra of 3C\,236. Data are taken from \cite{Schulz2018}. Fit parameters are listed in Table \ref{tab:Data:fitparam}. Top: VLA spectrum with three Gaussian components fitted to it (coloured dotted lines). Their combination is shown as the black dashed line. Bottom: Clipped integrated VLBI sepctra for the core and lobe region of 3C\,236}
			\label{fig:3C236_spec_fit}%
		\end{figure}
		
		\begin{figure}[h]
			\centering
			\includegraphics[width=0.97\linewidth]{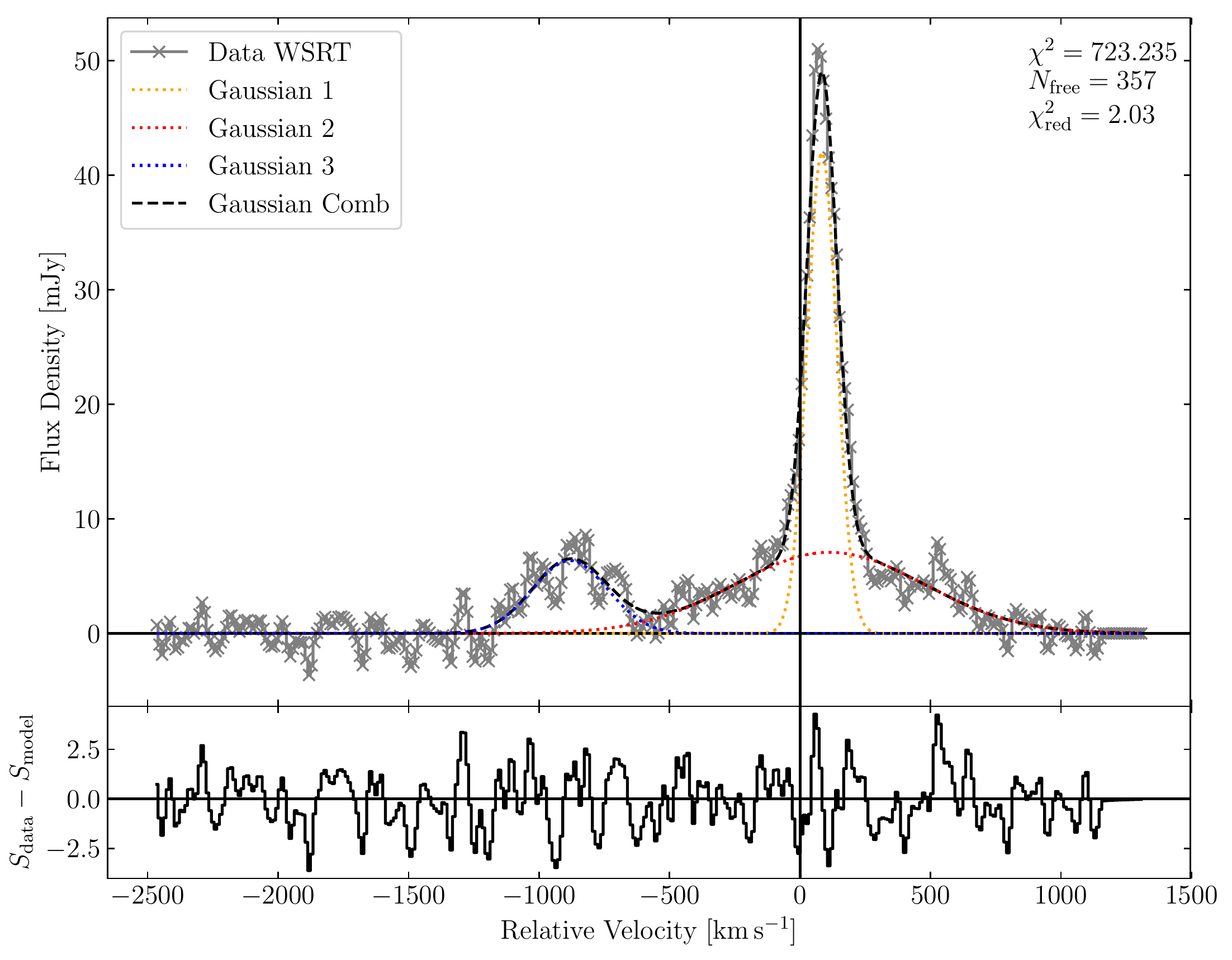}
			\caption{Inverted WSRT \ion{H}{I} absorption spectrum of 4C12.50 from \cite{Morganti2013}. Three Gaussian components were fitted to it (colored dotted lines) and their combination is shown as the black dashed line. Fit parameters are listed in Table \ref{tab:Data:fitparam}}
			\label{fig:4C12_spec_fit}%
		\end{figure}

\end{appendix}

\end{document}